\documentclass[letter]{aa} 
\usepackage{graphicx}
\usepackage{amsmath, mathtools}
\usepackage{txfonts}
\newcommand{\sub }[1]{_{\mathrm{#1}}}
\newcommand{\Mstar}{M_{\star}}
\newcommand{\Lstar}{L_{\star}}

\newcommand{\Teff}{T\sub{eff}}
\newcommand{\Mdot}{\dot{M}\sub{acc}}
\newcommand{\Mseed}{M\sub{seed}}
\newcommand{\Zseed}{Z\sub{seed}}
\newcommand{\Msun}{  M_\odot }
\newcommand{\Rsun}{  R_\odot }
\newcommand{\Lsun}{  L_\odot}
\newcommand{\tsun}{  t_\odot }
\newcommand{\Mearth}{ M_\oplus}

 \newcommand{ \Ys }{ Y\sub{surf} }
 \newcommand{ \Zs }{ Z\sub{surf} }
 \newcommand{ \ZXs }{ (Z/X)\sub{surf} }
 \newcommand{ \Zc }{ Z\sub{center} }
 \newcommand{ \Xc }{ X\sub{center} }
 
 \newcommand{ \Tc }{ T\sub{center} }
 \newcommand{ \kapc }{ \kappa\sub{center} }
 \newcommand{ \Xacc }{ X\sub{accretion} } 
 \newcommand{ \Yacc }{ Y\sub{accretion} } 
 \newcommand{ \Zacc }{ Z\sub{accretion} }

 \newcommand{ \Zaccini }{ Z\sub{proto} }
 \newcommand{ \Zaccmax }{ Z\sub{acc,max} }
 \newcommand{ \delcs }{ \delta c_s }
 \newcommand{ \RCZ }{ R\sub{CZ} }
 \newcommand{ \Mlost }{M_{XY,\rm{lost}}}
 \newcommand{ \Mpl }{M_{Z,\rm{planet}}}
 \newcommand{ \Xproto }{X\sub{proto}}
 \newcommand{ \Yproto }{Y\sub{proto}}
 \newcommand{ \Zproto }{Z\sub{proto}}
 \newcommand{ \chitwo }{ \chi^2 }
\newcommand{ \phiB}{ \varPhi(\element[][8]{B}) }
\newcommand{ \phiBe}{ \varPhi(\element[][7]{Be}) }
\newcommand{ \phiN}{ \varPhi(\element[][13]{N}) }
\newcommand{ \phiO}{ \varPhi(\element[][15]{O}) }
\newcommand{ \phiF}{ \varPhi(\element[][17]{F}) }
\newcommand{ \phipp}{ \varPhi( {pp} ) }
\newcommand{ \phipep}{ \varPhi( {pep} ) }
\newcommand{ \phiCNO}{ \varPhi( \mathrm{CNO} ) }

\defcitealias{GS98}{GS98}
\defcitealias{Asplund+09}{AGSS09}

\usepackage[colorlinks = true,
        linkcolor = blue,
        urlcolor  = blue,
        citecolor = blue,
        anchorcolor = blue
]{hyperref}
\makeatletter
\renewcommand*\aa@pageof{, page \thepage{} of \pageref*{LastPage}}
\makeatother

\bibpunct{(}{)}{;}{a}{}{,} 

\begin{document}

\title{Evidence of a signature of planet formation processes from solar neutrino fluxes\thanks{
            Supplemental materials are available at the CDS via anonymous ftp to cdsarc.u-strasbg.fr (***.**.***.*) via \url{http://cdsarc.u-strasbg.fr/viz-bin/qcat?J/A+A/***/***}; or via \url{https://doi.org/10.5281/zenodo.7156794}
      }}

\author{Masanobu Kunitomo\inst{1}
\and
Tristan Guillot\inst{2}
\and
Ga\"{e}l Buldgen\inst{3}
}

\institute{Department of Physics, Kurume University, 67 Asahimachi, Kurume, Fukuoka 830-0011, Japan\label{inst1}\\
\email{kunitomo.masanobu@gmail.com}
\and
Universit\'e C\^ote d'Azur, Observatoire de la C\^ote d'Azur, CNRS, Laboratoire Lagrange, Bd de l’Observatoire, CS 34229, 06304 Nice cedex 4, France\label{inst2}
\and
D\'{e}partement d'Astronomie, Universit\'{e} de Gen\`{e}ve, Chemin Pegasi 51, CH-1290 Versoix, Switzerland\label{inst3}
}
\date{Received 2 June 2022 / Accepted 7 October 2022}
\authorrunning{M. Kunitomo, T. Guillot \& G. Buldgen}


\abstract
{
      Solar evolutionary models are thus far unable to reproduce spectroscopic, helioseismic, and neutrino constraints consistently, resulting in the so-called solar modeling problem. In parallel, planet formation models predict that the evolving composition of the protosolar disk and, thus, of the gas accreted by the proto-Sun must have been variable. We show that solar evolutionary models that include a realistic planet formation scenario lead to an increased core metallicity of up to $5\%$, implying that accurate neutrino flux measurements are sensitive to the initial stages of the formation of the Solar System.
      Models with homogeneous accretion match neutrino constraints to no better than $2.7\sigma$. In contrast, accretion with a variable composition due to planet formation processes, leading to metal-poor accretion of the last $\sim$4\% of the young Sun's total mass, yields solar models within $1.3\sigma$ of all neutrino constraints. We thus demonstrate that in addition to increased opacities at the base of the convective envelope, the formation history of the Solar System constitutes a key element in resolving the current crisis of solar models.
}



\maketitle
%


\section{Introduction}
\label{sec:intro}

The Sun is a key piece of the puzzle of the theory of stellar structure and evolution. Due to its proximity, a great wealth of observational data has been acquired (i.e., spectroscopic, helioseismic, and neutrino observations). All these constraints have been extensively used to model our star theoretically using numerical simulations. Good agreement was achieved in the 1990s using so-called standard solar models \citep[solar models constructed with standard input physics; e.g.,][]{Christensen-Dalsgaard+96}.
In recent years, updates of the solar chemical composition \citep{Asplund+09, Asplund+21} have significantly worsened the situation. This so-called solar abundance problem has been thoroughly studied in the last two decades, with a clear solution yet to be found \citep{Montalban2006,Basu2008,Buldgen+19b,Orebi-Gann+21,Magg+22}.
While the abundance revision triggered the crisis, the issue is actually more of a solar modeling problem, with multiple ingredients called into question.

Apart from abundances, the main suspect has been radiative opacity, which over the course of the history of stellar evolution theory has been known to require significant revisions.
A number of studies identified that an increase of $\sim$10\%  in the interior opacity would indeed partly alleviate the issue \citep{Bahcall+05, Ayukov+Baturin17, Kunitomo+Guillot21}.
Recently, experimental measurements of iron opacity in almost the same conditions as the solar interior \citep{Bailey+15} have lent further traction to this hypothesis, even though these results still have to be explained from a theoretical point of view \citep[see, amongst others,][]{Nahar2016,Iglesias2017}.

In addition to microscopic ingredients such as opacity, the recipe of the standard solar model itself is under question. Recently, \citet{Eggenberger+22} showed that including the effects of (magneto)hydrodynamic instabilities in evolutionary computations allows us to simultaneously reproduce the solar internal rotation profile, the surface lithium abundance, and the surface helium abundance, a feat impossible for standard solar models that neglect rotation. This refinement provides only a partial solution to the issue, as it does not improve the agreement of models in terms of sound speed inversions and neutrino fluxes.

The first detection of solar neutrinos was surprising because inferred fluxes were about three times smaller than predicted by standard solar models. This so-called solar neutrino problem was an issue due to a change in the flavor of the neutrinos themselves and is now resolved \citep[e.g.,][and references therein]{Christensen-Dalsgaard21}.
Recent experimental studies have measured neutrinos both from the proton-proton ($pp$) chain and the carbon-nitrogen-oxygen (CNO) cycle, and the accuracy has been improving \citep[e.g.,][]{Agostini+18, Borexino-Collaboration20, Orebi-Gann+21}.
Solar neutrinos are most important because they provide direct access to the solar core: $pp$, $^7$Be, and $^8$B neutrinos provide information on the temperature and temperature gradient in the core, whereas CNO neutrinos provide direct access to the core composition \citep{Haxton+Serenelli+08, Gough19}.
Although many studies have attempted to account for measured solar neutrino fluxes \citep[e.g.,][]{Bahcall+06}, no solar model that uses updated abundances has reproduced both neutrino observations and other constraints simultaneously \citep[e.g.,][]{Vinyoles+17}.
Matching the observational constraints globally is therefore paramount for progress in our understanding of the composition, evolution, and early history of the Sun.

In this study we combine solar evolutionary models that include the early accretion phase with up-to-date knowledge of planet formation processes to calculate present-day solar models. These models are then compared to all available observational constraints.

\begin{figure*}[ht]
      \begin{center}
            \includegraphics[angle=0,width=0.8\hsize]{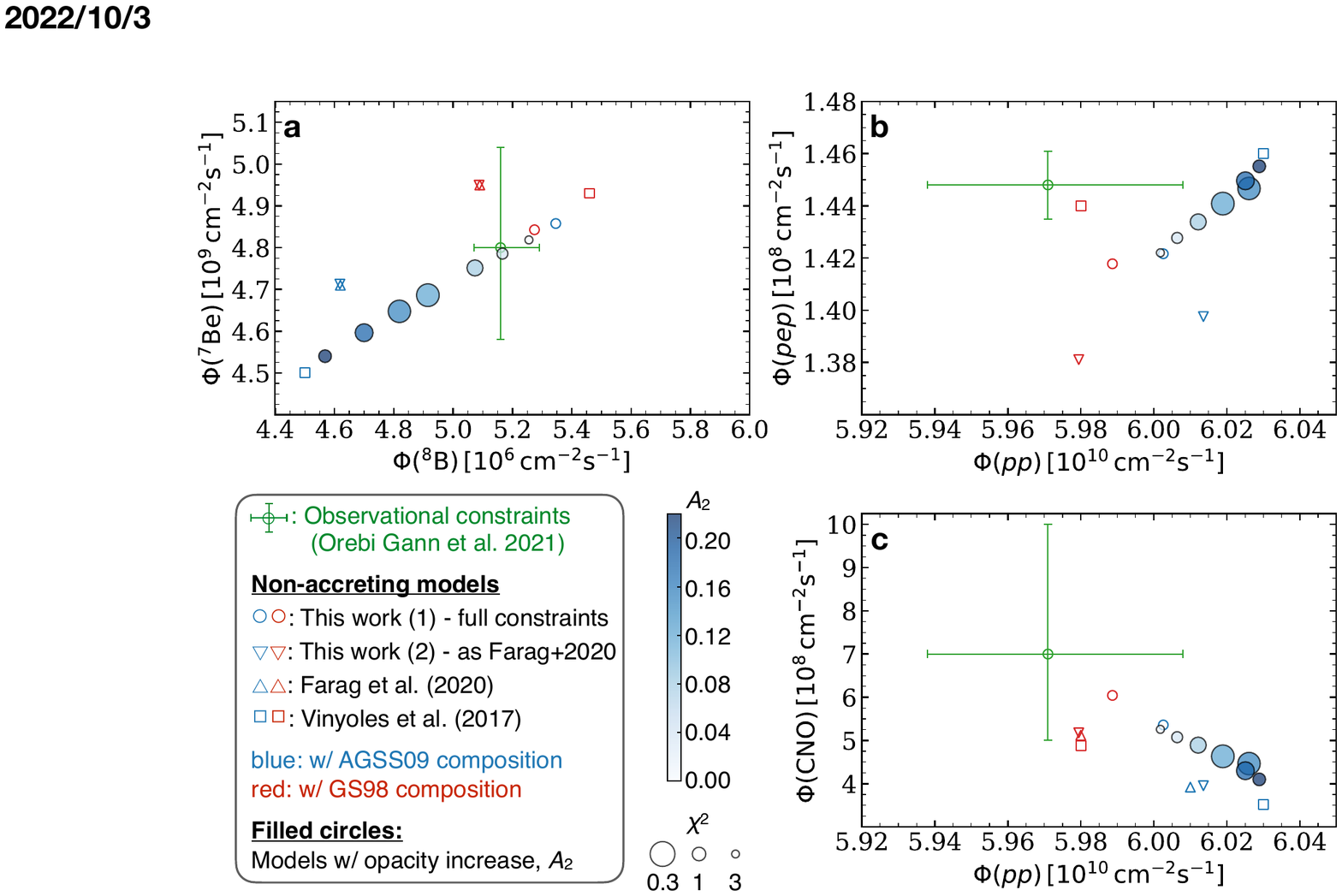}
      \end{center}
      \caption{Neutrino fluxes obtained for standard solar models and models with opacity increases.
            Panels ({\bf a}), ({\bf b}), and ({\bf c}) show the neutrino fluxes of $\phiBe$, $\phiB$, $\phipp$, $\phipep$, and $\phiCNO=\phiN+\phiO+\phiF$ (see Sect.\,\ref{sec:nuc}).
            The squares and the upward and downward pointing triangles show the standard solar models with three calibration constraints (see Sect.\,\ref{sec:calib}), those from \cite{Vinyoles+17}, \cite{Farag+20}, and this study, respectively.
            The open circles show our standard solar models with six constraints.
            The $\phipep$ and $\phiF$ values are not available in \cite{Farag+20}, and thus $\phiCNO=\phiN+\phiO$ is used for their models.
            The blue and red colors indicate the models with low-$Z$ \citetalias{Asplund+09} and high-$Z$ \citetalias{GS98} compositions, respectively.
            The filled circles show our models with opacity increase $A_2$ ($=0$, 0.04, 0.08, 0.12, 0.15, 0.18, and 0.22), which is shown by the color, with the size $\chitwo$, which is $\lesssim$0.5 for $A_2\in[0.12, 0.18]$.
            The green circles with error bars show the observed constraints \citep{Orebi-Gann+21}.
      }
      \label{fig:noacc}
\end{figure*}

\section{Calibrating solar models that include disk accretion} \label{sec:model}

We focused on solar models calibrated to match both helioseismic and spectroscopic constraints using an extended procedure (see Sect.\,\ref{sec:calib} and Table\,\ref{tab:results}) and examined how they fare against the observed neutrino fluxes  (Sect.\,\ref{sec:nuc}). We followed the approach of our previous work \citep{Kunitomo+Guillot21}, where we used solar models computed with the Modules for Experiments in Stellar Astrophysics (\texttt{MESA}) stellar evolution code, taking the progressive accretion of circumstellar disk material onto the proto-Sun into account  (Sect.\,\ref{sec:calc}).
We adopted the spectroscopic abundances from \citet[][\citetalias{Asplund+09} hereafter]{Asplund+09}; this choice did not have an effect on our conclusions.
Our extended calibration procedure aims to reproduce six constraints: luminosity, effective temperature, surface helium abundance, the surface abundance ratio of metals to hydrogen, location of the convective-radiative boundary, and relative differences in the sound speed profile.
We calibrated three different sets of models, refining the input physics each time. First, standard solar models were considered. Second, we considered nonstandard models, including an opacity increase near the base of the convective envelope, diffusive overshooting, and accretion with a homogeneous composition. Finally, accretion with a variable composition due to planet formation processes was considered.

Several studies \citep{Guzik+05, Castro+07, Haxton+Serenelli+08, Serenelli+11, Zhang+19} have indicated that a variation in the accretion metallicity, $\Zacc$, would affect the outcome of the calibration. On the basis of models of the evolution of circumstellar disks and of the formation and migration of pebbles, planetesimals, and protoplanets \citep{Garaud07}, we advocate for the existence of a rapid drift of pebbles (``pebble wave'') followed by a decrease in the metallicity of the accreted gas due  to both the early accretion of this dust and the formation of planetesimals and protoplanets \citep{Kunitomo+Guillot21}.
Disk winds, either thermally or magnetically driven, may also selectively remove hydrogen and helium and thus limit the extent and duration of the low-metallicity accretion phase.
A comparison of the mass of dust measured in protoplanetary disks and the mass of solids in exoplanets \citep{Manara+18} suggests that dust growth and the formation of planetesimals must have started early, within a million years of the formation of the central protostars. In our Solar System, isotopic anomalies in meteorites are indicative of the formation of Jupiter's core $\sim$1\,Myr after the first solids \citep{Kruijer+2020}, indicating that an efficient filtering of incoming dust \citep{Guillot+14} and low-metallicity accretion should have begun at that time. We estimate that $97$ to $168\,\Mearth$ of solids should be retained in planets or ejected from the Solar System \citep{Kunitomo+18}.
This corresponds to the accretion of $\sim$2 to $4\%$ of the Sun's mass in the form of zero-metallicity gas. The presence of a pebble wave leads to an early accretion of solids and thus an increase in this value. Conversely, the selective removal of hydrogen and helium by disk winds lowers this value.
Consequently, estimates based on up-to-date simulations of the evolution of disks indicate that between 2 and 5\% of the mass last accreted by the Sun should be characterized by a low metallicity \citep[][]{Kunitomo+Guillot21}.
In our best model, the accretion metallicity increases from $0.014$ (protosolar mass $\leq$0.90\,$\Msun$ and time $\leq$0.7\,Myr) to $0.065$ ($0.96\,\Msun$ and $2.2$\,Myr),
$0.1\,\Msun$ metal-free gas is lost by disk winds,
and $150\,\Mearth$ of solids are retained in planets,
leading to a $0.04\,\Msun$ metal-free gas in the last phase (see Fig.\,\ref{fig:schematic}b and Sect.\,\ref{sec:planet} for a detailed description).

\section{Results} \label{sec:results}

\begin{figure*}[ht]
      \begin{center}
            \includegraphics[angle=0,width=0.8\hsize]{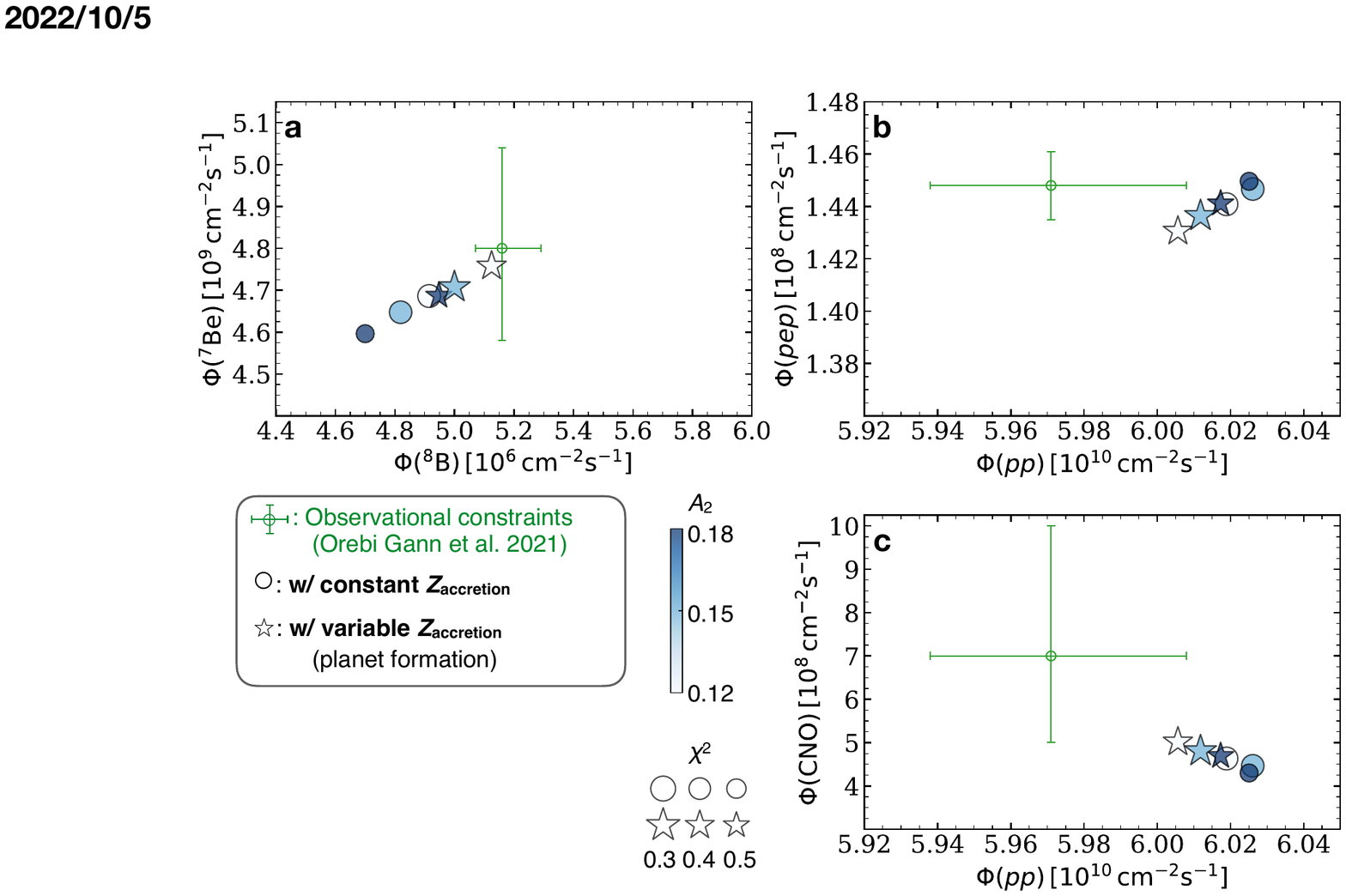}
      \end{center}
      \caption{
            Neutrino fluxes obtained for the models with an opacity increase and planet formation.
            Shown are the models with an opacity increase (circles; see also Fig.\,\ref{fig:noacc}) and with both an opacity increase and a variable $\Zacc$ (i.e., planet formation processes; star symbols).
            The colors show the opacity increase $A_2\in[0.12, 0.18]$. The opacity increase in this range leads to a better match with spectroscopic and helioseismic observations ($\chitwo\lesssim0.5$ indicated by the size).
            A higher opacity increase leads to lower $\phiB$, $\phiBe$, and $\phiCNO$ values, whereas the planet formation processes lead to higher values. Consequently, our best model with both an opacity increase of $A_2=0.12$ and planet formation processes (i.e., star symbol with white color) reproduces not only the observed constraints of neutrino fluxes (green circles with error bars) but also the spectroscopic and helioseismic observations.
            (See ``K2-MZvar-A2-12'' and ``K2-A2-12'' in Table\,\ref{tab:results} and the online table for more details about the models with $A_2=0.12$ and with and without planet formation processes, respectively.)
      }
      \label{fig:planet}
\end{figure*}

\begin{figure*}[ht]
      \begin{center}
            \includegraphics[angle=0,width=\hsize]{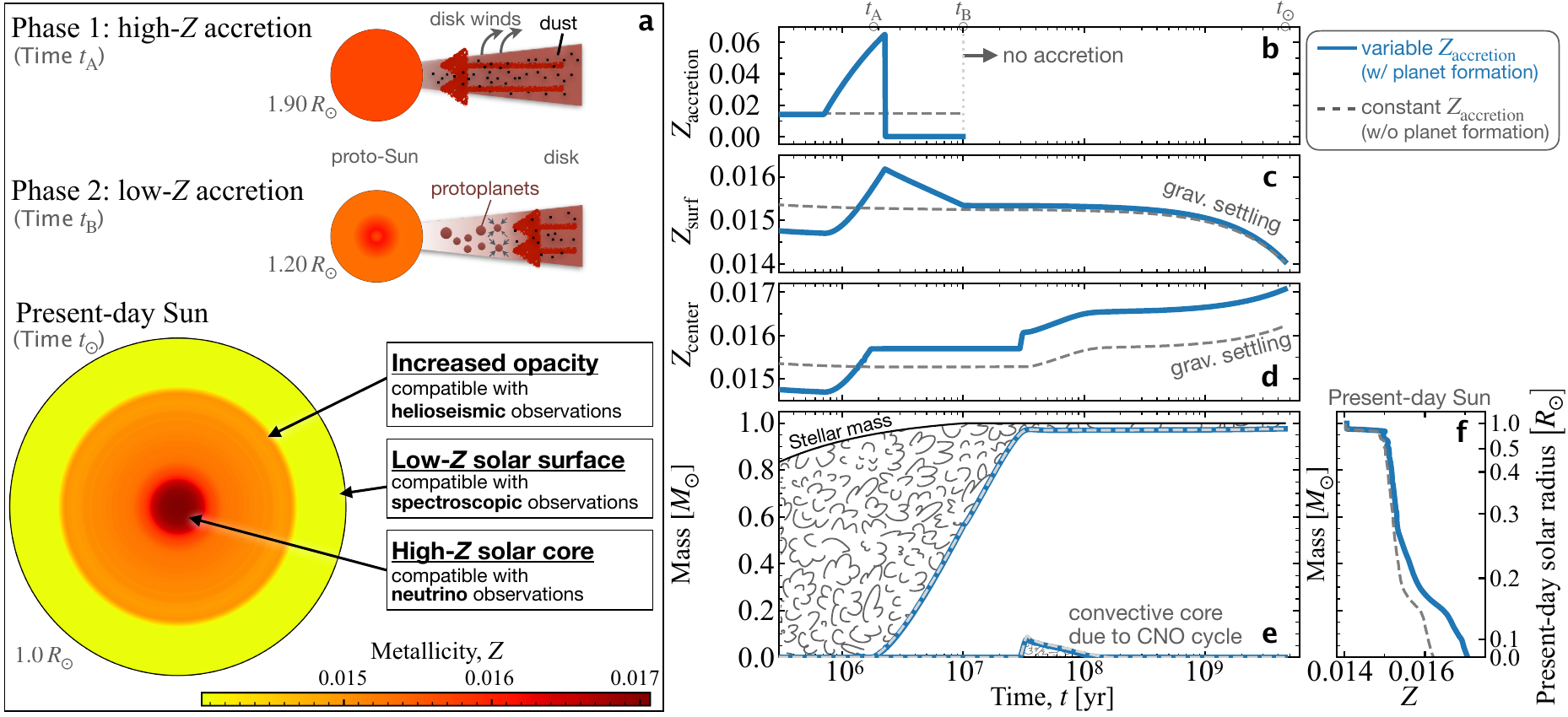}
      \end{center}
      \caption{
            {
                        Long-term evolution of models with planet formation processes.
                        Panel ({\bf a}) is a schematic illustration of the evolution of a model with a variable $\Zacc$ and an opacity increase of $A_2=0.12$. In the early phase, the metallicity of the proto-Sun increases with time due to the high-$Z$ accretion, which results from the dust drift and selective removal of hydrogen and helium by disk winds (see text). Once protoplanets are formed or dust grains in the protosolar disk are exhausted, the disk metallicity decreases, and consequently, the solar surface metallicity, $Z\sub{surf}$, decreases. The signature of this variable $\Zacc$ remains in the solar core until the solar age. The circles (not to scale) show the solar metallicity profile at times
                        $t\sub{A}=1.73\,$Myr,
                        $t\sub{B}=10\,$Myr, and
                        $\tsun=4.567$\,Gyr.
                        Panels ({\bf b}), ({\bf c}), and ({\bf d}) show the evolution of $\Zacc$, $\Zs$, and $\Zc$ of the models with a variable $\Zacc$ (solid blue lines) and a constant $\Zacc$ (dotted gray lines), respectively. Panel ({\bf e}) shows the internal structure evolution (the so-called Kippenhahn diagram). The cloudy and white regions show the convective and radiative regions, respectively. The two lines (blue and gray) are almost the same. A radiative core emerges at time $t\sub{A}$. The solid black line indicates the stellar mass. Panel ({\bf f}) is the metallicity profile of the present-day Sun.
                        Animations that show the metallicity profile in the solar interior of these models are available at \href{https://doi.org/10.5281/zenodo.7156794}{Zenodo}.
                        (See ``K2-MZvar-A2-12'' and ``K2-A2-12'' in Table\,\ref{tab:results} and the online table for more details regarding the models with and without planet formation processes, respectively.)
                  }
      }
      \label{fig:schematic}
\end{figure*}

In Fig.\,\ref{fig:noacc} we compare the neutrino fluxes ($\phiBe$, $\phiB$, $\phipp$, $\phipep$, and $\phiCNO$; see Sect.\,\ref{sec:nuc}) from observational constraints to those obtained for standard solar models and for models that include an opacity increase.  Standard solar models in the literature \citep{Vinyoles+17, Farag+20} indicate higher $\phiBe$, $\phiB$, and $\phiCNO$ for models with high-metallicity ($Z$) \citet[][\citetalias{GS98} hereafter]{GS98} abundances compared to models with low-$Z$ \citetalias{Asplund+09} abundances.
Our calculations show that this is in fact a calibration issue, linked to the fact that these standard solar models used three constraints as usual in a classical calibration scheme (see Sect.\,\ref{sec:calib}): when using the same approach with the three calibration constraints, we obtain the same results as \citet{Farag+20} (see our Fig.\,\ref{fig:noacc}). Conversely, when including our extended calibration procedure with six constraints \citep[similar to][]{Ayukov+Baturin17}, the results obtained for \citetalias{GS98} and \citetalias{Asplund+09} abundances are found to be close to one another and good matches to the neutrino constraints. However, these models with the \citetalias{Asplund+09} composition are poor fits to the spectroscopic and helioseismic constraints \citep{Kunitomo+Guillot21}, with values of $\chitwo$ significantly above unity, illustrating the well-known disagreement between the standard solar models with the \citetalias{Asplund+09} abundances and the observations.

Models that include an opacity increase (parameter $A_2$ corresponds to the relative increase to the standard opacity), shown in Fig.\,\ref{fig:noacc}, are characterized by lower $\chitwo$ values \citep{Bahcall+05, Ayukov+Baturin17, Kunitomo+Guillot21}, but they depart from the observed neutrino fluxes due to their low central metallicities. Our best models that minimize $\chitwo$ (with $A_2=0.12$ to $0.18$) have a $\phiB$ flux that is $2.7$ to $5.1\sigma$ lower than the observational constraints, whereas both $\phipp$ and $\phiCNO$ are within 1.5 and 1.4$\sigma$, respectively. Therefore, although these models match both helioseismic and spectroscopic constraints, they do not account for the observed neutrino fluxes. This is a well-known issue with solar models that led some works to consider low-metallicity accretion to try to recover a ``high-$Z$'' radiative interior embedded under a ``low-$Z$'' convective envelope \citep{Guzik+05,Castro+07}.

Thus far, we have only considered solar models that either start with the Sun's present mass or include the accretion of disk material with a constant metallicity.
Figure\,\ref{fig:planet} compares the models with the lowest values of $\chitwo$ in Fig.\,\ref{fig:noacc} (i.e., those with an opacity increase of $A_2 \in [0.12, 0.18]$ and homogeneous accretion) with the same models but including accretion with a variable composition.
The values of $\chitwo$ are almost independent of our choice of the accretion model: the planet formation processes occur while the proto-Sun is still largely convective, implying that most of the solar internal structure is unchanged. However, the values of all neutrino fluxes are affected, indicating that the solar core retains a clear signature of planet formation processes. Values of  $\phiB$, $\phiBe$, and $\phiCNO$ increase, whereas $\phipp$ and $\phipep$ decrease slightly \citep{Serenelli+11}. This results in models with planet formation processes that fit all the helioseismic, spectroscopic, and neutrino constraints. This is the case for the model with $A_2=0.12$, with a worse fit being $\phipep$, which is only
$1.3\sigma$ lower than the observational constraint.

The enhancement of the \element[][7]{Be}, \element[][8]{B}, and CNO neutrino fluxes is directly linked to an increased metallicity and temperature of the solar core, which are closely related (see Sect.\,\ref{sec:nuc}).
As shown in Fig.\,\ref{fig:schematic}, the pebble wave leads to an initially high metallicity throughout the solar interior (see Figs.\,\ref{fig:schematic}a and \ref{fig:schematic}b). The following low-$Z$ accretion phase leads to a decrease in the surface metallicity, which, because of the spectroscopic constraints included in the calibration, becomes indistinguishable from that in the case with homogeneous accretion (Fig.\,\ref{fig:schematic}c). However, the central region retains memory from the initial high-$Z$ phase because the growth of the central radiative core after 1.7\,Myr halts its mixing with the rest of the star (Figs.\,\ref{fig:schematic}d and \ref{fig:schematic}e). Between 30 and 100\,Myr, an incomplete CNO cycle leads to a transient central convective core (also due to out-of-equilibrium \element[][3]{He} burning) and to a limited increase in the central metallicity. It then progressively increases on timescales of $\sim$1\,Gyr due to gravitational settling. This results in a central metallicity of the present-day Sun that is increased by $\simeq$5\% over that of models that do not account for planet formation processes (see Figs.\,\ref{fig:schematic}f and \ref{fig:Zevol}).

We stress that the magnitude of this effect cannot be precisely estimated due to uncertainties in both star and planet formation processes. The duration of the accretion phase may be shorter than assumed here (10\,Myr), leading to a smaller increase in the central metallicity \citep{Kunitomo+Guillot21}. On the other hand, the release of energy by the accretion shock on the proto-Sun could lead to a faster growth of the internal radiative zone after only $\sim$1\,Myr \citep{Kunitomo+18}. Similarly, including mass loss by solar winds leads to a higher mass of the proto-Sun and hence also leads to a faster growth of the radiative zone. Thus, both processes are expected to amplify the change in the central metallicity and neutrino fluxes due to the planet formation processes, regardless of the adopted solar chemical composition.

\section{Conclusion} \label{sec:conclusion}

We simulated the evolution of the Sun from the protostellar phase to the present day using all available constraints. Our study demonstrates that planet formation processes have a significant contribution to present-day solar neutrino fluxes.
The main conclusions are summarized as follows:
\begin{itemize}
      \item The solution depends on the calibration procedure: our models calibrated with six helioseismic and spectroscopic constraints are different from standard solar models calibrated with three constraints.
      \item Models that include a localized opacity increase, as indicated by experimental results, match all the helioseismic and spectroscopic constraints. However, due to their low central temperature, they match the neutrino constraints to no better than $2.7\sigma$.
      \item The variable composition of accreted gas resulting from planet formation processes induces a higher central metallicity of the present-day Sun by up to 5\%, which increases $\phiB$, $\phiBe$, and $\phiCNO$ and lowers $\phipp$ and $\phipep$, in agreement with observational requirements. Our best model reproduces all the observed neutrino fluxes within 1.3$\sigma$.
\end{itemize}

In conclusion, despite the uncertainties related to planet formation scenarios, our work shows that neutrino fluxes are intrinsically tied to the evolutionary history of the Sun and that the effects of planet formation should be considered when interpreting these measurements.
In our best models, pebble waves and the formation of planets led to a metal-poor accretion of the last 4\% of the young Sun's mass.
More accurate neutrino fluxes measurements and constraints on the input physics that affect neutrino fluxes are thus highly desirable.

\begin{acknowledgements}
      This work was supported by the JSPS KAKENHI (grant no. 20K14542).
      Numerical computations were carried out on the PC cluster at the Center for Computational Astrophysics, National Astronomical Observatory of Japan. T.G. acknowledges funding from the {\it Programme National de Plan\'{e}tologie}. G.B. acknowledges funding from the SNF AMBIZIONE grant no. 185805 (Seismic inversions and modelling of transport processes in stars).
      \textit{Software}: \texttt{MESA} \citep[version 12115;][]{Paxton+11,Paxton+13,Paxton+15,Paxton+18, Paxton+19}.
\end{acknowledgements}

%
%
\bibliographystyle{aa}

\begin{appendix} 

      \section{Methods}
      \subsection{Calculation of solar structure and evolution models}\label{sec:calc}
      This section describes how we simulated the formation and evolution of the Sun. We followed Sect.\,3 of \cite{Kunitomo+Guillot21} for the modeling unless otherwise noted.
      For more details, we refer readers to \cite{Kunitomo+17}, \cite{Kunitomo+18}, and \cite{Paxton+11, Paxton+13, Paxton+15, Paxton+18, Paxton+19}.
      Our models were calculated using the \texttt{MESA} stellar evolution code \citep{Paxton+11, Paxton+13, Paxton+15, Paxton+18, Paxton+19}.
      The models of nuclear reactions and accretion are discussed below.
      We modeled the opacity increase as a function of temperature. Since the contribution of iron to the Rosseland-mean opacity has a Gaussian form with temperature \citep{LePennec+15}, we adopted the Gaussian function centered at $T=10^{6.45}\,$K with varying amplitude, $A_2$ \citep[see Eq.\,5 of][]{Kunitomo+Guillot21}.
      Convection is modeled with the mixing-length theory from \cite{Cox+Giuli68}. Diffusive convective overshooting is also considered with the model in \cite{Herwig00} both below a convective envelope and above a convective core.
      Element diffusion is calculated with the formalism in \cite{Thoul+94}.

      There are four differences from standard solar models \citep[e.g.,][]{Vinyoles+17,Farag+20}: namely, the calibration procedure, diffusive overshooting, opacity increase, and accretion that has a variable composition.
      We included planet formation processes based on recent progress, which differs from previous studies, including low-metallicity accretion (see below).
      We note that although \cite{Eggenberger+22} recently showed the impact of (magneto)hydrodynamic instabilities driven by rotation on the internal rotation and lithium abundance, they do not improve sound speed inversions and neutrino fluxes, and thus, we do not account for these instabilities.

      \subsection{Calibration procedure}\label{sec:calib}

      In each model, input parameters are calibrated to minimize the reduced $\chitwo$ value defined as $\chi^2 = {\sum_{i=1}^N \left[  (q_i-q_{i,\rm target})/\sigma(q_i)\right]^2}/{N}$, where $N$ is the number of constraints, $q_i$ is the simulation result, and $q_{i,\rm target}$ and $\sigma$ are the observational constraint and its uncertainty, respectively.
      We adopted $N=6$ for the models with an ``extended'' calibration procedure \citep[similar to][]{Ayukov+Baturin17} and $N=3$ (i.e., $\ZXs$, $\Lstar$, $\Teff$; see below) for the standard solar models.
      Using the downhill simplex method \citep{Nelder+Mead65}, we searched for a set of input parameters that minimizes $\chitwo$. We assumed that the solar age is 4.567\,Gyr \citep{Amelin+02}.
      The input parameters include two parameters for convection (mixing length and overshooting) and parameters for the initial composition, opacity increase, and accretion, which vary from model to model.

      In the calibration procedure, simulation results were compared with the six constraints from spectroscopic and helioseismic observations summarized in Table\,\ref{tab:results} \citep[see also Table\,3 of][]{Kunitomo+Guillot21}:
      surface abundance ratio of metals to hydrogen $\ZXs$,
      surface helium abundance $\Ys$,
      location of the convective-radiative boundary $\RCZ$,
      root-mean-square sound speed rms($\delcs$),
      luminosity $\Lstar$ ($=\Lsun\pm0.01\,$dex),
      and
      effective temperature $\Teff$ ($=5777\pm10$\,K).
      We note that the current uncertainty in the surface metallicity of the Sun \citep{Asplund+21,Magg+22} does not affect the conclusions of our study; a thorough discussion is beyond the scope of this work, which instead focuses on the effect of planet formation processes.

      \begin{table*}[ht]
            \begin{center}
                  \caption{Results minimized by the chi-squared simulations and observational constraints.}
                  \label{tab:results}
                  \scalebox{0.9}[0.9]{
                        \begin{tabular}{lllllllllllll} 
                              \hline\hline
                              Model name                                           &
                              $\chitwo$                                            &
                              rms($\delcs$)                                        &
                              $\ZXs$                                               & $\Ys$       & $\RCZ$      &
                              $\varPhi(pp)$                                        &
                              $\varPhi(pep)$                                       &
                              $\phiBe$                                             &
                              $\phiB$                                              &
                              $\phiN$                                              &
                              $\phiO$                                              &
                              $\varPhi( \mathrm{CNO} )$                                                                                                                                                                                                                                       \\
                                                                                   &             & [\%]        &               &              & [$\Rsun$]    &
                              [$10^{10}$]                                          & [$10^{8}$]  & [$10^{9}$]  & [$10^{6}$]    &
                              [$10^{8}$]                                           & [$10^{8}$]  & [$10^{8}$]                                                                                                                                                                                 \\
                              \noalign{\smallskip}
                              \hline
                              \noalign{\smallskip}
                              \multicolumn{7}{l}{ [\textit{Previous studies}] }    &                                                                                                                                                                                                          \\
                              V+17-AGSS09\tablefootmark{a}                         & 4.72        & 0.2         & {\bf 0.0178}  & 0.232        & {\bf 0.722}  & 6.03                 & {\bf 1.46}  & {\bf 4.50}         & 4.50               & {\bf 2.04}  & {\bf 1.44}  & 3.48
                              \\
                              V+17-GS98\tablefootmark{a}                           & {\bf 0.52}  & {\bf 0.05}  & {\bf 0.0230}  & 0.243        & {\bf 0.712}  & {\bf 5.98}           & {\bf 1.44}  & {\bf 4.93}         & 5.46               & {\bf 2.78}  & {\bf 2.05}  & 4.83
                              \\
                              F+20-AGSS09\tablefootmark{b}                         & --          & N/A         & {\bf 0.0181}  & 0.240        & {\bf 0.726}  & 6.01                 & N/A         & {\bf 4.71}         & 4.62               & {\bf 2.25}  & {\bf 1.67}  & 3.92
                              \\
                              F+20-GS98\tablefootmark{b}                           & --          & N/A         & {\bf 0.0229}  & {\bf 0.246}  & {\bf 0.718}  & {\bf 5.98}           & N/A         & {\bf 4.95}         & {\bf 5.09}         & {\bf 2.91}  & {\bf 2.21}  & {\bf 5.12}
                              \\
                              \noalign{\smallskip}
                              \hline
                              \noalign{\smallskip}
                              \multicolumn{7}{l}{ [\textit{This study}] }          &                                                                                                                                                                                                          \\
                              SSM-AGSS09\tablefootmark{c}                          & 4.70        & 0.45        & {\bf 0.0181}  & 0.240        & {\bf 0.726}
                                                                                   & 6.013       & 1.398       & {\bf 4.713}   & 4.618        & {\bf 2.248}  & {\bf 1.672}          & 3.953                                                                                                   \\
                              SSM-GS98\tablefootmark{c}                            & {\bf 0.67}  & 0.18        & {\bf 0.0229}  & {\bf 0.246}  & {\bf 0.718}
                                                                                   & {\bf 5.979} & {\bf 1.381} & {\bf 4.949}   & {\bf 5.088}  & {\bf 2.914}  & {\bf 2.214}          & {\bf 5.176}                                                                                             \\
                              noacc-noov                                           & 2.81        & 0.32        & 0.0204        & {\bf  0.247} & 0.723
                                                                                   & {\bf 6.003} & 1.422       & {\bf 4.858}   & {5.346}      & {\bf 2.973}  & {\bf 2.342}          & {\bf 5.359}
                              \\
                              noacc-GS98-noov                                      & {\bf 0.62}  & 0.16        & {\bf  0.0239} & {\bf  0.248} & {\bf  0.717}
                                                                                   & {\bf 5.989} & 1.418       & {\bf 4.843}   & {\bf 5.274}  & {\bf 3.352}  & {\bf 2.636}          & {\bf 6.043}
                              \\
                              K2-A2-12                                             & {\bf 0.36}  & {\bf 0.099} & {\bf  0.0190} & {\bf  0.246} & {\bf  0.715}
                                                                                   & {6.019}     & {\bf 1.441} & {\bf 4.686}   & {4.915}      & {\bf 2.588}  & {\bf 2.008}          & {4.633}
                              \\
                              {\bf K2-MZvar-A2-12}                                 & {\bf 0.39}  & 0.11        & {\bf  0.0190} & {\bf  0.247} & {\bf  0.716} & {\bf 6.006}          & 1.431       & {\bf 4.758}        & {\bf 5.125}        & {\bf 2.785} & {\bf 2.188} & {\bf 5.014}
                              \\
                              \noalign{\smallskip}
                              \hline
                              \noalign{\smallskip}
                              \multicolumn{7}{l}{ [\textit{Adopted constraints}] } &                                                                                                                                                                                                          \\
                              Value                                                & --          & 0           & $0.0181$      & $0.2485$     & $0.713$      & $5.971$              & $1.448$     & $4.80$             & $5.16$             & $\leq 13.7$ & $\leq 2.8$  & $7.0^{+3.0}_{-2.0}$ \\
                              Uncertainty, $\sigma$                                & --          & 0.1         & $0.001$       & $0.0035$     & $0.01$       & $^{+0.037}_{-0.033}$ & $0.013$     & $^{+0.24}_{-0.22}$ & $^{+0.13}_{-0.09}$ &             &                                   \\

                                                                                   &             &             & $0.02292$     &              &              &                      &             &                    &                    &                                                 \\
                                                                                   &             &             & $0.001$                                                                                                                                                                      \\
                              References                                           & --          & 1           & 2, 3          & 4            & 5, 6         & 7, 8                 & 7, 8        & 7, 8               & 7, 8               & 9           & 9           & 9                   \\
                              \noalign{\smallskip}
                              \hline
                              \noalign{\smallskip}
                        \end{tabular}
                  }
            \end{center}
            \tablefoot{
            Neutrino fluxes on Earth, $\varPhi$, are in units of $\rm cm^{-2}s^{-1}$.
            The numbers highlighted in bold indicate the values that satisfy the constraints or have $\chi^2<1$.
            The K2-MZvar-A2-12 model is our best model with a variable accretion metallicity.
            Two constraints of $\ZXs$ are shown for the models with the \citetalias{Asplund+09} (top) and \citetalias{GS98} (bottom) compositions. See Sect.\,\ref{sec:calib} for the description of models.
            This table is available in electronic form at the CDS and \href{https://doi.org/10.5281/zenodo.7156794}{Zenodo}.
            \tablefoottext{a}{V+17 refers to \citet[][see their Tables 4 and 6]{Vinyoles+17}.
                  We derived the $\chitwo$ of their models assuming a perfect matching of the luminosity and effective temperature.}
            \tablefoottext{b}{F+20 refers to \citet[][see their Tables 1, 2, and 3]{Farag+20}. We note that rms($\delcs$), $\phipep$, and $\phiF$ are not available; thus, $\phiCNO=\phiN+\phiO$ for their models.}
            \tablefoottext{c}{Our simulations with the same settings as \cite{Farag+20}.}
            {\bf References.} (1) \cite{Basu+09}, (2) \citetalias{Asplund+09}, (3) \citetalias{GS98}, (4) \cite{Basu+Antia04}, (5) \cite{Bahcall+05}, (6) \cite{Kunitomo+Guillot21}, (7) \cite{Bergstrom+16}, (8) \cite{Orebi-Gann+21}, (9) \cite{Borexino-Collaboration20}.
            }
      \end{table*}

      Table\,\ref{tab:results} also summarizes the results of our models and previous studies.
      The SSM-AGSS09 and SSM-GS98 models are our standard solar models following \cite{Farag+20} (i.e., $N=3$) with the \citetalias{Asplund+09} and \citetalias{GS98} abundances, respectively. The noacc-noov and noacc-GS98-noov models (open circles in Figs.\,\ref{fig:noacc} and \ref{fig:M2}e) are the models with the extended calibration ($N=6$) without overshooting.
      The model with opacity increase ($A_2=0.12$) and overshooting is denoted K2-A2-12 (filled circles in Figs.\,\ref{fig:noacc} and \ref{fig:planet}), whereas the same model but with planet formation processes is denoted K2-MZvar-A2-12 (star symbols in Fig.\,\ref{fig:planet}).
      Additional supplemental materials showing the details of the results of the optimized models are available at the CDS and \href{https://doi.org/10.5281/zenodo.7156794}{Zenodo} (see the links provided in the footnote on the first page).

      We note that \citet[][]{Haxton+Serenelli+08} investigated the dependence of neutrino fluxes on parameters by using the results of standard solar models and derived their relations assuming the linearity of the dependence. In contrast, we performed evolutionary simulations with the iterative calibration procedure in order to self-consistently  track the evolution history, including the effects of planet formation processes in the early phase, and to obtain a global structure of the present-day solar interior model that matches all the spectroscopic, helioseismic, and neutrino constraints.

      \subsection{Nuclear reactions and neutrinos}\label{sec:nuc}
      Solar neutrinos are emitted by two types of thermonuclear reactions, namely the \textit{pp} chain and the CNO cycle.
      The \textit{pp} chain is
      \small
      \begin{align}
            \begin{rcases}
                  {{p(p, e^+\nu_1)d}}\\
                  {{p(e^-p,\nu_2)d}}
            \end{rcases}
            d(p,\gamma)\element[][3]{He}
            \begin{cases}
                  \element[][3]{He}(\element[][3]{He}, 2p)\element[][4]{He} \\ 
                  \element[][3]{He}(\element[][4]{He}, \gamma)\element[][7]{Be}
                  \begin{cases}
                        {{\element[][7]{Be} (e^-, \nu_3)}}\element[][7]{Li}(p,\element[][4]{He})\element[][4]{He} \\ 
                        \element[][7]{Be} (p, \gamma){{\element[][8]{B}(e^+\nu_4)}}\element[][8]{Be}(\element[][4]{He})\element[][4]{He}, 
                  \end{cases}
            \end{cases}
            \label{eq:pp}
      \end{align}
      \normalsize
      where \textit{p} is a proton and \textit{d} is a deuterium atom.
      Another known neutrino-emitting reaction is $\element[][3]{He}(p, e^+\nu)\element[][4]{He}$, but it is extremely rare in the current Sun \citep{Vinyoles+17,Agostini+18}.
      The fluxes of neutrinos $\nu_1$, $\nu_2$, $\nu_3$, and $\nu_4$ observed on Earth are denoted by $\phipp$, $\phipep$, $\phiBe$, and $\phiB$, respectively.
      The two branches of the CNO cycle are
      \begin{align}
             & \element[][12]{C}(p, \gamma){{\element[][13]{N}(e^+\nu_5)}}\element[][13]{C}(p, \gamma)\element[][14]{N}(p,\gamma){{\element[][15]{O}(e^+\nu_6)}}\element[][15]{N}(p,\element[][4]{He})\element[][12]{C}\,, \label{eq:CNO} \\
             & \element[][15]{N}(p, \gamma)\element[][16]{O}(p,\gamma){{\element[][17]{F}(e^+\nu_7)}}\element[][17]{O}(p,\element[][4]{He})\element[][14]{N}\,.
            \label{eq:CNO2}
      \end{align}
      The fluxes of $\nu_5$, $\nu_6$, and $\nu_7$ are $\phiN$, $\phiO$, and $\phiF$, respectively.
      We define $\phiCNO$ as $\phiN+\phiO+\phiF$ unless otherwise noted.

      Neutrino fluxes depend on abundances and temperature. The following fluxes are in particular strongly sensitive to temperature as
      $\phiBe\propto \Xc\Zc \Tc^{11}$,
      $\phiB\propto\Xc\Zc\Tc^{25}$,
      and $\phiN,\ \phiO\propto\Xc\Zc\Tc^{20}$
      \citep{Bahcall+Ulmer96}.
      In contrast, $\phipp$ is more sensitive to the hydrogen abundance and less sensitive to temperature as $\propto \Xc^2 \Tc^{4}$ \citep{Kippenhahn+Weigert90}.
      Therefore, in the models with a higher $\Zc$ and $\Tc$, $\phiBe$, $\phiB$, and $\phiCNO$ are higher, whereas such models have a lower hydrogen abundance $\Xc$ and thus a lower $\phipp$ to satisfy the constraint $\Lstar = \Lsun$ at the solar age (see Figs.\,\ref{fig:noacc} and \ref{fig:planet}).

      We note that $\Zc$ and $\Tc$ are closely related, as follows. A higher $\Zc$ leads to a higher central opacity, $\kapc$, because metals dominantly contribute to the opacity.
      With a higher $\kapc$, the temperature gradient needs to be higher to flow out the luminosity in the radiative core \citep{Kippenhahn+Weigert90}, and thus $\Tc$ is higher.

\begin{figure*}[ht]
            \begin{center}
                  \includegraphics[angle=0,width=0.7\hsize]{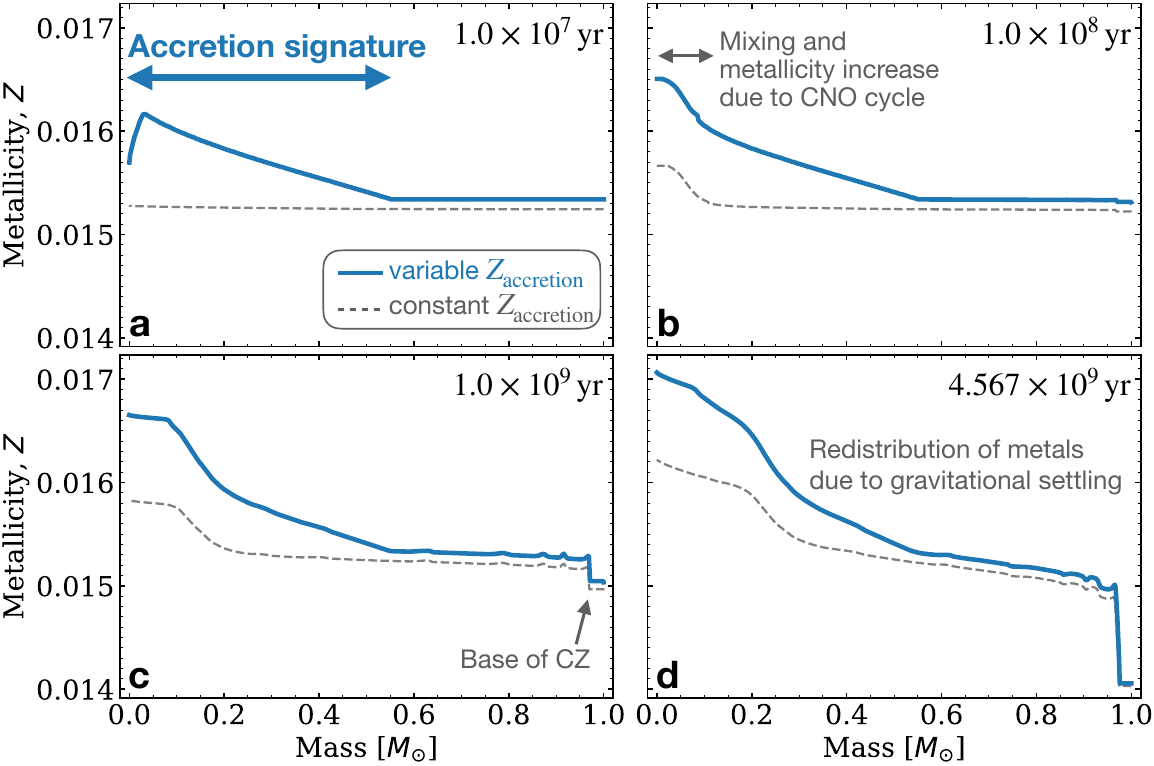}
            \end{center}
            \caption{
            Evolution of the metallicity profile with mass coordinate.
            The solid blue and dashed gray lines show the K2-MZvar-A2-12 and K2-A2-12 models, respectively, at 10\,Myr (end of the accretion phase; panel {\bf a}), 100\,Myr ({\bf b}), 1\,Gyr ({\bf c}), and 4.567\,Gyr (solar age; {\bf d}).
            Animations showing the metallicity profile in the solar interior of these models are available at \href{https://doi.org/10.5281/zenodo.7156794}{Zenodo}.
            }
            \label{fig:Zevol}
      \end{figure*}

      In the main branch of the CNO cycle (Eq.\,\ref{eq:CNO}), the bottleneck reaction is $\element[][14]{N}(p,\gamma)\element[][15]{O}$. Although the CNO cycle produces only a small fraction ($\approx$1\%) of energy in $1\,\Msun$ stars because of a slightly low $\Tc$, a tentative ``incomplete CNO cycle'' occurs from 30 to 100\,Myr. This leads to the conversion of preexisting $^{12}$C and $2p$ into $^{14}$N, and thus, the central metallicity, $\Zc$, is increased (see Eq.\,\ref{eq:CNO} and Fig.\,\ref{fig:schematic}d).
      The CNO cycle has another important implication: it leads to a sharp temperature gradient in the core and thus results in the presence of a convective core \citep{Iben65}. In our best model, the central region has a composition gradient in the pre-main-sequence phase, and thus, mixing in the core leads to an increase in metallicity at 30\,Myr (see Figs.\,\ref{fig:schematic}d and \ref{fig:Zevol}), whereas this $\Zc$ modification at 30\,Myr does not appear in our model with a homogeneous $\Zacc$.

      We adopted the nuclear reaction rates from NACRE \citep{Angulo+99} and \cite{Imbriani+05} for the reaction rate of $\element[][14]{N}(p,\gamma)\element[][15]{O}$.
      To output neutrino fluxes, we used the subroutine provided in \cite{Farag+20} with a minor update to output $\phiF$.
      The electron screening formalism of \cite{Chugunov+07} was used.
      We also mention that the neutrino fluxes are sensitive to the uncertainties in the nuclear reaction rates, as well as electron screening, opacity tables, and the transport of chemical elements over the course of solar evolution \citep{Boothroyd2003, Salmon+21, Vescovi+19, Villante+Serenelli21, Eggenberger+22}, and thus could be affected by potential future revisions of these ingredients.

\begin{figure}
            \begin{center}
                  \includegraphics[angle=0,width=\hsize]{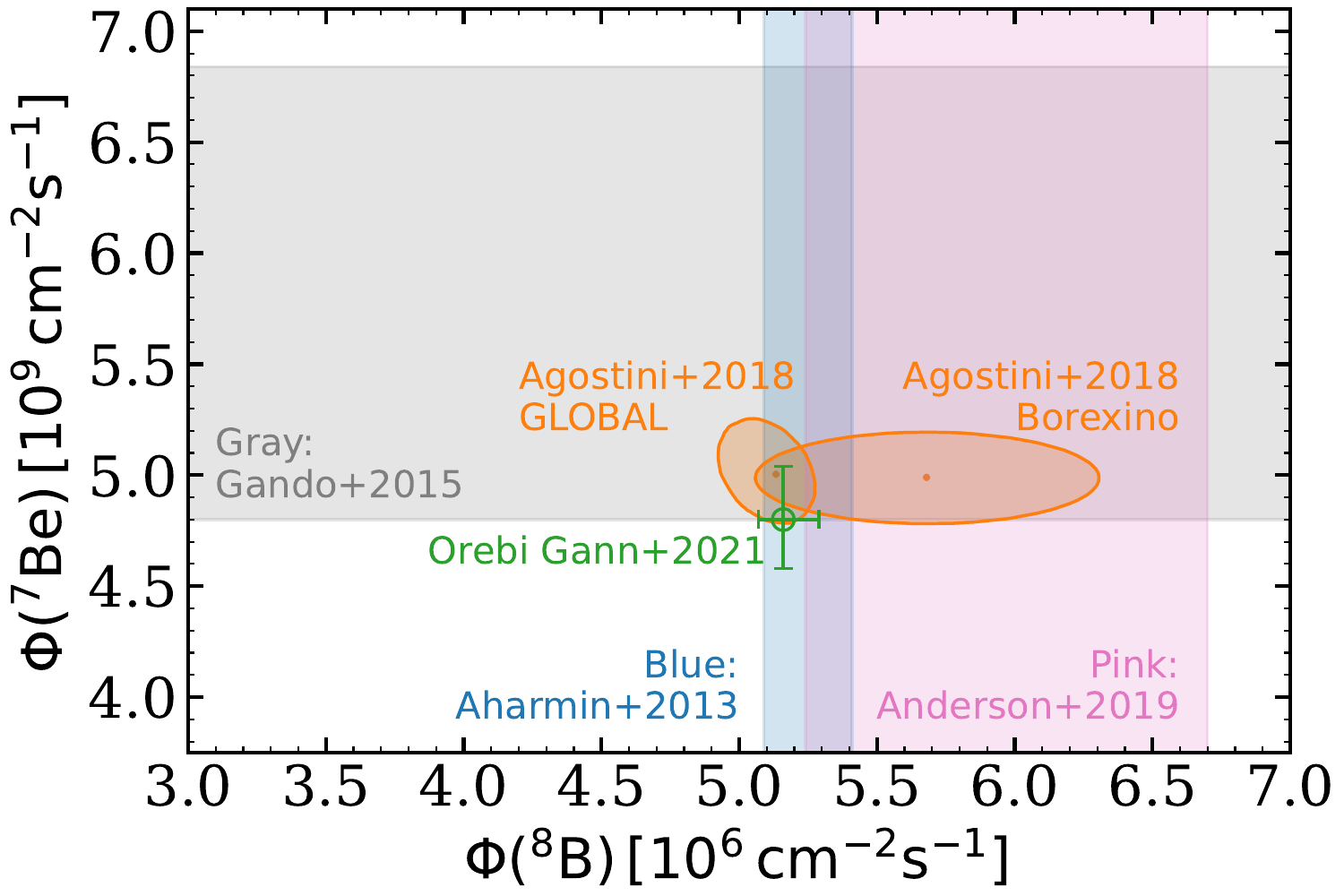}
            \end{center}
            \caption{
                  Observational constraints on the $\phiB$ and $\phiBe$ values.
                  The green open circle with error bars shows the constraints derived using all experimental neutrino data available up to 2016 \citep{Bergstrom+16,Orebi-Gann+21}, which is used in this study.
                  The orange ellipses show the constraints in \citet{Agostini+18} from the Borexino experiments only (right) and from using available experimental data (left).
                  The gray, blue, and pink shaded regions indicate the constraints from the KamLAND experiments \citep{Gando+15}, from \cite{Aharmim+13} with the Sudbury Neutrino Observatory (SNO), and from \cite{Anderson+19} with the SNO+, respectively.
            }
            \label{fig:obs}
      \end{figure}

      The calibrated models are compared with observed neutrino fluxes in Figs.\,\ref{fig:noacc} and \ref{fig:planet}.
      Much effort has been made in experiments around the world to constrain the solar neutrino fluxes.
      We adopted the $\phipp$, $\phipep$, $\phiB$, and $\phiBe$ values from \cite{Bergstrom+16} and \cite{Orebi-Gann+21}, which were derived using all the data available up to 2016, whereas we adopted the recent constraint in \cite{Borexino-Collaboration20} for $\phiCNO$. We note that there remain relatively large uncertainties in the observed fluxes (partially originated from the uncertainties in the neutrino oscillation model) and that a disagreement between various works is still being debated \citep[][see Fig.\,\ref{fig:obs}]{Zyla+20, Orebi-Gann+21}.\ However, discussing these uncertainties is beyond the scope of our work. Nevertheless, future experiments to constrain the fluxes more precisely are thus highly encouraged.

      \subsection{Planet formation processes}\label{sec:planet}
      We followed \cite{Kunitomo+Guillot21} for accretion processes onto the proto-Sun: we started with calculations from a protostellar phase with a seed of mass $\Mseed=0.1\,\Msun$ and metallicity $\Zseed=0.02$.
      We adopted the accretion rate, $\Mdot$, based on observations \citep{Hartmann+98}: it decreases with time and stops at 10\,Myr. We considered heat injection by accreted materials \citep{Kunitomo+17,Kunitomo+Guillot21}.
      We did not consider mass loss by solar winds in the main-sequence phase \citep{Guzik1987, Sackmann+Boothroyd03, Wood+05, Zhang+19}.

      In the models with planet formation processes, the composition of the accreted materials ($\Xacc$, $\Yacc$, and $\Zacc$ for the mass fractions of hydrogen, helium, and metals, respectively) vary with time.
      In the models without them, we adopted a constant composition with time as $\Xacc=\Xproto$, $\Yacc=\Yproto$, and $\Zacc=\Zproto$.

      We modeled planet formation processes as a time-dependent $\Zacc$ based on recent planet formation theory following \cite{Kunitomo+Guillot21} (see also Fig.\,\ref{fig:schematic}b).
      In the earliest phase, the disk gas has a primordial composition (i.e., $\Xproto$, $\Yproto$, and $\Zproto$). Then, we considered a high $\Zacc$ value due to dust drift \citep{Appelgren+20, Elbakyan+20} and/or disk winds \citep{Guillot+Hueso06, Kunitomo+20}. In the last phase, $\Zacc=0$ due to the depletion or filtration of pebbles by a planetary gap \citep{Guillot+14, Sato+16}.
      In the models with planet formation, three input parameters are calibrated in addition to those for the primordial composition and convection: namely, the stellar masses when $\Zacc$ increases and decreases ($M_1$ and $M_2$, respectively) and the maximum value of $\Zacc$, $\Zaccmax$.

      We note that $\Zacc$ must be consistent with the observations of the Solar System planets and planet formation theory. The total mass of metals in the Solar System planets, $\Mpl$, is estimated to be between 97 and 168\,$\Mearth$ \citep{Kunitomo+18}. Recent studies have suggested that dust grains grow in the protostellar phase \citep{Manara+18,Tsukamoto+17} and a proto-Jupiter was formed early, $\sim$1\,Myr after the Ca-Al-rich inclusions (CAIs) condensed \citep{Kruijer+2020}, implying low $M_1$ and $M_2$ values. The total mass lost by disk winds, $\Mlost$, is still uncertain, but a recent study \citep{Kunitomo+20} suggested that it can be $\sim$0.1\,$\Msun$ around a $1\,\Msun$ star.
      If we neglect the early high-$Z$ accretion phase (pebble wave) and assume $\Zproto\approx0.015$, this leads to a $\approx$0.03\,$\Msun$ low-$Z$ accretion in the late phase (i.e., $M_2\approx0.97\,\Msun$). Under the fixed value of $\Zproto$, the pebble wave leads to a lower $M_2$, whereas metal-poor disk winds lead to a higher $M_2$.
      In our best model (see ``K2-MZvar-A2-12'' in Table\,\ref{tab:results} and the online table), $\Zproto=0.014$,
      $\Zaccmax=0.065$, $M_1=0.90\,\Msun$, $M_2=0.96\,\Msun$, $\Mlost=0.10\,\Msun$, and $\Mpl=150\,\Mearth$.

      Figure\,\ref{fig:Zevol} shows how the signature of planet formation processes remains in the metallicity profile of the present-day Sun (especially $\lesssim$0.2\,$\Msun$). At the end of the accretion phase ($=$10\,Myr), a nonhomogeneous metallicity profile due to the time-dependent $\Zacc$ exists in the radiative core ($\lesssim$0.5\,$\Msun$). In particular, high-$Z$ accretion is imprinted below $0.05\,\Msun$. The central region ($\lesssim$0.1\,$\Msun$) is subject to an incomplete CNO cycle between 30 and 100\,Myr that leads to the convective mixing and a metallicity increase of $\sim$5$\times10^{-4}$. Gravitational settling takes effect on a timescale of $\sim$1\,Gyr leading to an increase in the central metallicity of $\sim$5$\times10^{-4}$. Therefore, the central composition of the present-day Sun is determined by accretion, convective mixing, nuclear reactions, and element diffusion.

\begin{figure}
            \begin{center}
                  \includegraphics[angle=0,width=\hsize]{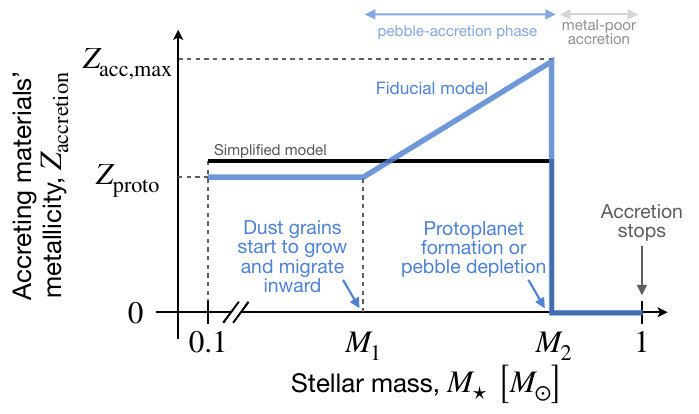}
            \end{center}
            \caption{
                  Time evolution of the metallicity of accreted gas. The blue line indicates our fiducial $\Zacc$ model, where it increases from $\Zproto$ at $\Mstar=M_1$ to $\Zaccmax$ at $M_2$ followed by a metal-poor accretion phase. The black line indicates the simplified model (model M2var; see Fig.\,\ref{fig:M2}), which has the same $M_2$ and the total amount of accreted metals as the blue line. If this $M_2$ value is lower than $0.96\,\Msun$ (which approximately corresponds to when the radiative core develops), then these two models lead to the same $\Zc$.
            }
            \label{fig:Zacc}
      \end{figure}

      We note that $\Mlost$, $\Mpl$, $\Zproto$, and $\Zaccmax$ are linked by mass conservation \citep{Kunitomo+Guillot21}:
      \begin{eqnarray}\label{eq:Mlost}
            \Zaccini(1\,\Msun+\Mlost) \!&=&\! \Zaccini M_2 \\
            \!&&\!+\frac12 (\Zaccmax-\Zproto) (M_2-M_1)  + \Mpl\,.\nonumber
      \end{eqnarray}
      In the early phase in which the proto-Sun is fully convective (i.e., when the protosolar mass $\Mstar< 0.95\,\Msun$), the accretion history is lost. This implies that solutions are highly degenerate: an increase in the metallicity due to a pebble wave that occurs during that phase is indistinguishable from the accretion of gas with a constant metallicity, as long as the total amount of heavy elements accreted over the period is identical (see Fig.\,\ref{fig:Zacc}).
      For example, our best model ($\Mlost=0.1\,\Msun$, $\Zaccmax=0.065$, $\Zproto=0.014$) is approximately equivalent to a simplified model with $M_1=M_2=0.96\,\Msun$, $\Zproto=0.016$, and $\Mlost\simeq 0$.
      Therefore, a large variety of solutions that lead to an increased central metallicity and neutrino fluxes agreeing with observations are possible.

\begin{figure*}
            \begin{center}
                  \includegraphics[angle=0,width=\hsize]{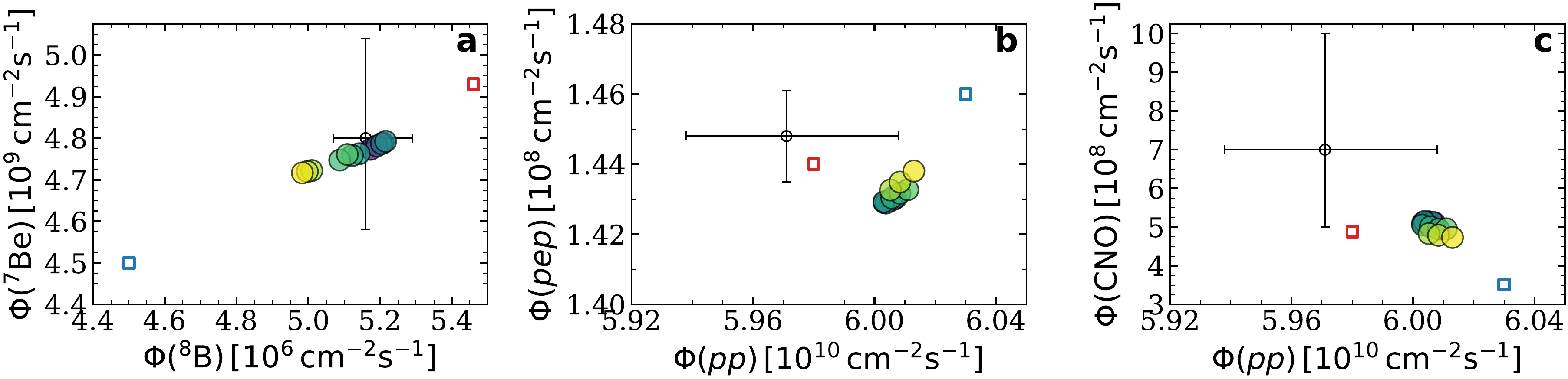}
                  \includegraphics[angle=0,width=0.42\hsize]{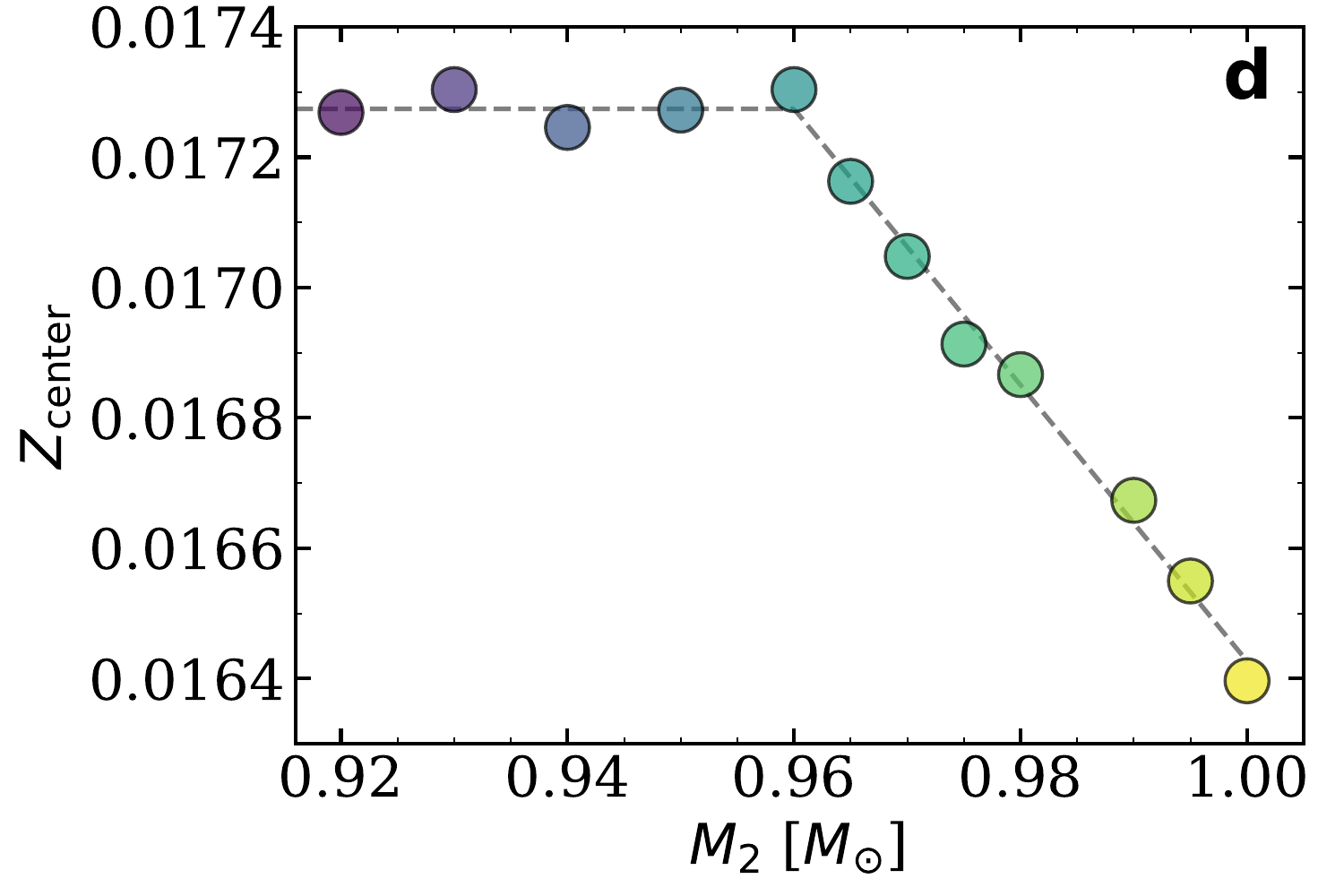}
                  \includegraphics[angle=0,width=0.42\hsize]{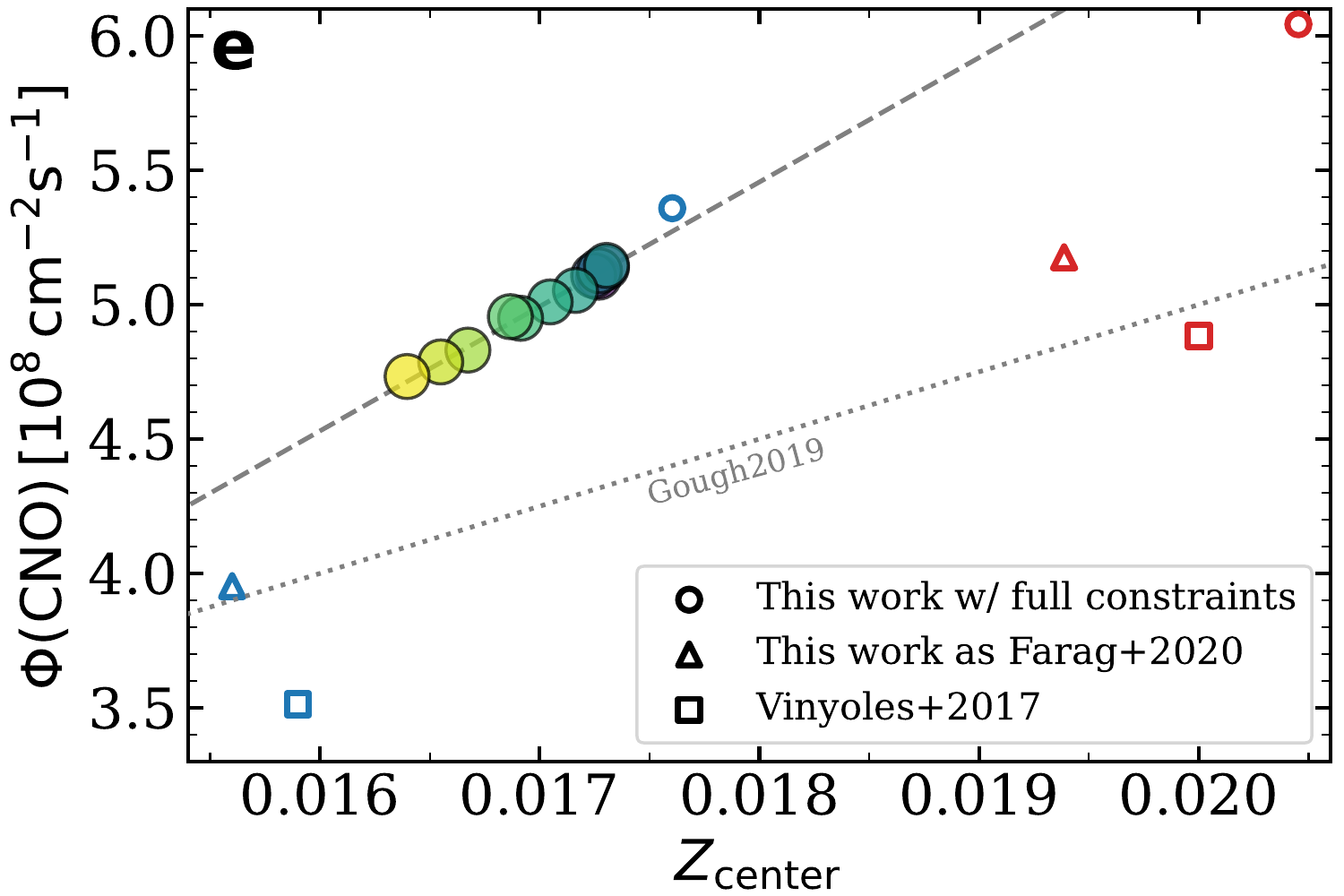}
            \end{center}
            \caption{
                  Neutrino fluxes and central metallicity of the models with different planet formation scenarios. The color of the filled circles (model M2var) corresponds to the $M_2$ value (see panel ({\bf d})), which is the protosolar mass when the accretion metallicity decreases. The dotted line in panel ({\bf e}) shows the relation in \cite{Gough19}, and the dashed lines in panels ({\bf d}) and ({\bf e}) show the fits of our M2var models.
                  The black circles with error bars show the observed constraints.
                  The red and blue points indicate the models with the old high-$Z$ \citetalias{GS98} composition and the low-$Z$ \citetalias{Asplund+09} composition, respectively.
            }
            \label{fig:M2}
      \end{figure*}

      To quantify the magnitude of the increased $\Zc$ by planet formation processes, we performed additional simulations (model ``M2var'') for various $M_2$ values. For simplicity, we neglected the high-$Z$ accretion phase (i.e., $\Zacc=\Zproto$ for $\Mstar<M_2$ and $\Zacc=0$ for $\Mstar>M_2$) and investigated the dependence of $\Zc$ on $M_2\in[0.92\,\Msun, 1\,\Msun]$.
      In addition, we neglected overshooting and fixed $A_2=0.12$. Figure\,\ref{fig:M2} shows that a lower $M_2$, which corresponds to a higher mass retained in planets, leads to a higher $\Zc$ and higher fluxes of $\phiB$, $\phiBe$, and $\phiCNO$. However, this effect saturates at $M_2=0.96\,\Msun$. This mass approximately corresponds to the timing when a radiative core develops (1.7\,Myr).
      In the models with a lower  $M_2<0.96\,\Msun$, a higher $\Zproto$ value is needed to reproduce the $\ZXs$ constraint, and thus, the total metal mass in the proto-Sun at 1.7\,Myr does not depend on $M_2$. Therefore, the formation of giant planets in the early phase ($\lesssim$2\,Myr), suggested by recent studies, is preferable to induce a high $\Zc$ value and explain the observed neutrino fluxes. From a linear regression analysis, we obtain the empirical relation
      \begin{eqnarray}
            \Zc = \min \left[0.01727,\ 0.01727- 0.0212 \left( M_2/\Msun - 0.960 \right) \right]\,.
      \end{eqnarray}

      In our M2var model, $\phiCNO$ has a positive correlation with $\Zc$. Recently, \cite{Gough19} suggested a relation $\phiCNO/(10^{8}\mathrm{cm^{-2}s^{-1}})= 250\, \Zc$ based on standard solar models \citep{Vinyoles+17}.
      Our models follow a different relation ($\phiCNO/(10^{8}\mathrm{cm^{-2}s^{-1}})= 463\, \Zc - 2.88$; see Fig.\,\ref{fig:M2}e) due to the different calibration procedures and the accretion history, which are not taken into account in the evolution of standard solar models.

\end{appendix}


\begin{thebibliography}{72}
      \expandafter\ifx\csname natexlab\endcsname\relax\def\natexlab#1{#1}\fi

      \bibitem[{Agostini {et~al.}(2018)Agostini, Altenm{\"u}ller, Appel,
                        Atroshchenko, Bagdasarian, Basilico, Bellini, Benziger, Bick, Bonfini, Bravo,
                        Caccianiga, Calaprice, Caminata, Caprioli, Carlini, Cavalcante, Chepurnov,
                        Choi, Collica, D'Angelo, Davini, Derbin, Ding, Di~Ludovico, Di~Noto,
                        Drachnev, Fomenko, Formozov, Franco, Gabriele, Galbiati, Ghiano, Giammarchi,
                        Goretti, Gromov, Guffanti, Hagner, Houdy, Hungerford, Ianni, Ianni, Jany,
                        Jeschke, Kobychev, Korablev, Korga, Kryn, Laubenstein, Litvinovich, Lombardi,
                        Lombardi, Ludhova, Lukyanchenko, Lukyanchenko, Machulin, Manuzio, Marcocci,
                        Martyn, Meroni, Meyer, Miramonti, Misiaszek, Muratova, Neumair, Oberauer,
                        Opitz, Orekhov, Ortica, Pallavicini, Papp, Penek, Pilipenko, Pocar, Porcelli,
                        Raikov, Ranucci, Razeto, Re, Redchuk, Romani, Roncin, Rossi, Sch{\"o}nert,
                        Semenov, Skorokhvatov, Smirnov, Sotnikov, Stokes, Suvorov, Tartaglia,
                        Testera, Thurn, Toropova, Unzhakov, Villante, Vishneva, Vogelaar, von
                        Feilitzsch, Wang, Weinz, Wojcik, Wurm, Yokley, Zaimidoroga, Zavatarelli,
                        Zuber, Zuzel, \& Collaboration}]{Agostini+18}
      Agostini, M., Altenm{\"u}ller, K., Appel, S., {et~al.} 2018, Nature, 562, 505

      \bibitem[{{Aharmim} {et~al.}(2013){Aharmim}, {Ahmed}, {Anthony}, {Barros},
      {Beier}, {Bellerive}, {Beltran}, {Bergevin}, {Biller}, {Boudjemline},
      {Boulay}, {Cai}, {Chan}, {Chauhan}, {Chen}, {Cleveland}, {Cox}, {Dai},
      {Deng}, {Detwiler}, {DiMarco}, {Doe}, {Doucas}, {Drouin}, {Duncan},
      {Dunford}, {Earle}, {Elliott}, {Evans}, {Ewan}, {Farine}, {Fergani},
      {Fleurot}, {Ford}, {Formaggio}, {Gagnon}, {Goon}, {Graham}, {Guillian},
      {Habib}, {Hahn}, {Hallin}, {Hallman}, {Harvey}, {Hazama}, {Heintzelman},
      {Heise}, {Helmer}, {Hime}, {Howard}, {Huang}, {Jagam}, {Jamieson}, {Jelley},
      {Jerkins}, {Keeter}, {Klein}, {Kormos}, {Kos}, {Kraus}, {Krauss}, {Kruger},
      {Kutter}, {Kyba}, {Lange}, {Law}, {Lawson}, {Lesko}, {Leslie}, {Loach},
      {MacLellan}, {Majerus}, {Mak}, {Maneira}, {Martin}, {McCauley}, {McDonald},
      {McGee}, {Miller}, {Monreal}, {Monroe}, {Nickel}, {Noble}, {O'Keeffe},
      {Oblath}, {Ollerhead}, {Orebi Gann}, {Oser}, {Ott}, {Peeters}, {Poon},
      {Prior}, {Reitzner}, {Rielage}, {Robertson}, {Robertson}, {Rosten},
      {Schwendener}, {Secrest}, {Seibert}, {Simard}, {Simpson}, {Skensved},
      {Sonley}, {Stonehill}, {Te{\v{s}}i{\'c}}, {Tolich}, {Tsui}, {Van Berg},
      {VanDevender}, {Virtue}, {Wan Chan Tseung}, {Wark}, {Watson}, {Wendland},
      {West}, {Wilkerson}, {Wilson}, {Wouters}, {Wright}, {Yeh}, {Zhang}, \&
      {Zuber}}]{Aharmim+13}
      {Aharmim}, B., {Ahmed}, S.~N., {Anthony}, A.~E., {et~al.} 2013, \prc, 88,
      025501

      \bibitem[{{Amelin} {et~al.}(2002){Amelin}, {Krot}, {Hutcheon}, \&
                        {Ulyanov}}]{Amelin+02}
      {Amelin}, Y., {Krot}, A.~N., {Hutcheon}, I.~D., \& {Ulyanov}, A.~A. 2002,
      Science, 297, 1678

      \bibitem[{{Anderson} {et~al.}(2019){Anderson}, {Andringa}, {Asahi}, {Askins},
      {Auty}, {Barros}, {Bartlett}, {Bar{\~a}o}, {Bayes}, {Beier}, {Bialek},
      {Biller}, {Blucher}, {Bonventre}, {Boulay}, {Caden}, {Callaghan}, {Caravaca},
      {Chauhan}, {Chen}, {Chkvorets}, {Cleveland}, {Connors}, {Coulter}, {Depatie},
      {Di Lodovico}, {Duncan}, {Dunger}, {Falk}, {Fischer}, {Fletcher}, {Ford},
      {Gagnon}, {Gilje}, {Grant}, {Grove}, {Hallin}, {Hallman}, {Hans}, {Hartnell},
      {Heintzelman}, {Helmer}, {Hern{\'a}ndez-Hern{\'a}ndez}, {Hreljac}, {Hu},
      {In{\'a}cio}, {Jillings}, {Kaptanoglu}, {Khaghani}, {Klein}, {Knapik},
      {Kormos}, {Krar}, {Kraus}, {Krauss}, {Kroupova}, {Lam}, {Land}, {Lane},
      {LaTorre}, {Lawson}, {Lebanowski}, {Leming}, {Li}, {Lidgard}, {Liggins},
      {Liu}, {Lozza}, {Luo}, {Maguire}, {Maio}, {Manecki}, {Maneira}, {Martin},
      {Marzec}, {Mastbaum}, {McCauley}, {McDonald}, {Mekarski}, {Meyer}, {Mlejnek},
      {Morton-Blake}, {Nae}, {Nirkko}, {O'Keeffe}, {Orebi Gann}, {Parnell},
      {Paton}, {Peeters}, {Pershing}, {Pickard}, {Pracsovics}, {Prior}, {Reichold},
      {Richardson}, {Rigan}, {Rose}, {Rosero}, {Rumleskie}, {Semenec}, {Singh},
      {Skensved}, {Stringer}, {Svoboda}, {Tam}, {Tian}, {Tseng}, {Turner}, {Van
                  Berg}, {Veinot}, {Virtue}, {V{\'a}zquez-J{\'a}uregui}, {Wang}, {Weigand},
      {Wilson}, {Woosaree}, {Wright}, {Yanez}, {Yeh}, {Zuber}, {Zummo}, \& {SNO +
                  Collaboration}}]{Anderson+19}
      {Anderson}, M., {Andringa}, S., {Asahi}, S., {et~al.} 2019, \prd, 99, 012012

      \bibitem[{{Angulo} {et~al.}(1999){Angulo}, {Arnould}, {Rayet}, {Descouvemont},
                        {Baye}, {Leclercq-Willain}, {Coc}, {Barhoumi}, {Aguer}, {Rolfs}, {Kunz},
                        {Hammer}, {Mayer}, {Paradellis}, {Kossionides}, {Chronidou}, {Spyrou},
                        {degl'Innocenti}, {Fiorentini}, {Ricci}, {Zavatarelli}, {Providencia},
                        {Wolters}, {Soares}, {Grama}, {Rahighi}, {Shotter}, \& {Lamehi
                                    Rachti}}]{Angulo+99}
      {Angulo}, C., {Arnould}, M., {Rayet}, M., {et~al.} 1999, Nuclear Physics A,
      656, 3

      \bibitem[{{Appelgren} {et~al.}(2020){Appelgren}, {Lambrechts}, \&
                        {Johansen}}]{Appelgren+20}
      {Appelgren}, J., {Lambrechts}, M., \& {Johansen}, A. 2020, \aap, 638, A156

      \bibitem[{{Asplund} {et~al.}(2021){Asplund}, {Amarsi}, \&
                        {Grevesse}}]{Asplund+21}
      {Asplund}, M., {Amarsi}, A.~M., \& {Grevesse}, N. 2021, \aap, 653, A141

      \bibitem[{{Asplund} {et~al.}(2009){Asplund}, {Grevesse}, {Sauval}, \&
                        {Scott}}]{Asplund+09}
      {Asplund}, M., {Grevesse}, N., {Sauval}, A.~J., \& {Scott}, P. 2009, \araa, 47,
      481

      \bibitem[{{Ayukov} \& {Baturin}(2017)}]{Ayukov+Baturin17}
      {Ayukov}, S.~V. \& {Baturin}, V.~A. 2017, Astronomy Reports, 61, 901

      \bibitem[{{Bahcall} {et~al.}(2005){Bahcall}, {Basu}, {Pinsonneault}, \&
                        {Serenelli}}]{Bahcall+05}
      {Bahcall}, J.~N., {Basu}, S., {Pinsonneault}, M., \& {Serenelli}, A.~M. 2005,
      \apj, 618, 1049

      \bibitem[{{Bahcall} {et~al.}(2006){Bahcall}, {Serenelli}, \&
                        {Basu}}]{Bahcall+06}
      {Bahcall}, J.~N., {Serenelli}, A.~M., \& {Basu}, S. 2006, \apjs, 165, 400

      \bibitem[{{Bahcall} \& {Ulmer}(1996)}]{Bahcall+Ulmer96}
      {Bahcall}, J.~N. \& {Ulmer}, A. 1996, \prd, 53, 4202

      \bibitem[{Bailey {et~al.}(2015)Bailey, Nagayama, Loisel, Rochau, Blancard,
                        Colgan, Cosse, Faussurier, Fontes, Gilleron, {et~al.}}]{Bailey+15}
      Bailey, J.~E., Nagayama, T., Loisel, G.~P., {et~al.} 2015, Nature, 517, 56

      \bibitem[{{Basu} \& {Antia}(2004)}]{Basu+Antia04}
      {Basu}, S. \& {Antia}, H.~M. 2004, \apjl, 606, L85

      \bibitem[{{Basu} \& {Antia}(2008)}]{Basu2008}
      {Basu}, S. \& {Antia}, H.~M. 2008, Physics Reports, 457, 217

      \bibitem[{{Basu} {et~al.}(2009){Basu}, {Chaplin}, {Elsworth}, {New}, \&
                        {Serenelli}}]{Basu+09}
      {Basu}, S., {Chaplin}, W.~J., {Elsworth}, Y., {New}, R., \& {Serenelli}, A.~M.
      2009, \apj, 699, 1403

      \bibitem[{{Bergstr{\"o}m} {et~al.}(2016){Bergstr{\"o}m}, {Gonzalez-Garcia},
                        {Maltoni}, {Pe{\~n}a-Garay}, {Serenelli}, \& {Song}}]{Bergstrom+16}
      {Bergstr{\"o}m}, J., {Gonzalez-Garcia}, M.~C., {Maltoni}, M., {et~al.} 2016,
      Journal of High Energy Physics, 2016, 132

      \bibitem[{{Boothroyd} \& {Sackmann}(2003)}]{Boothroyd2003}
      {Boothroyd}, A.~I. \& {Sackmann}, I.~J. 2003, \apj, 583, 1004

      \bibitem[{{Borexino Collaboration} {et~al.}(2020){Borexino Collaboration},
                        {Altenm{\"u}ller}, {Appel}, {Atroshchenko}, {Bagdasarian}, {Basilico},
                        {Bellini}, {Benziger}, {Biondi}, {Bravo}, {Caccianiga}, {Calaprice},
                        {Caminata}, {Cavalcante}, {Chepurnov}, {D'Angelo}, {Davini}, {Derbin}, {Di
                                    Giacinto}, {Di Marcello}, {Ding}, {Di Ludovico}, {Di Noto}, {Drachnev},
                        {Formozov}, {Franco}, {Galbiati}, {Ghiano}, {Giammarchi}, {Goretti},
                        {G{\"o}ttel}, {Gromov}, {Guffanti}, {Ianni}, {Ianni}, {Jany}, {Jeschke},
                        {Kobychev}, {Korga}, {Kumaran}, {Laubenstein}, {Litvinovich}, {Lombardi},
                        {Lomskaya}, {Ludhova}, {Lukyanchenko}, {Lukyanchenko}, {Machulin}, {Martyn},
                        {Meroni}, {Meyer}, {Miramonti}, {Misiaszek}, {Muratova}, {Neumair},
                        {Nieslony}, {Nugmanov}, {Oberauer}, {Orekhov}, {Ortica}, {Pallavicini},
                        {Papp}, {Pelicci}, {Penek}, {Pietrofaccia}, {Pilipenko}, {Pocar}, {Raikov},
                        {Ranalli}, {Ranucci}, {Razeto}, {Re}, {Redchuk}, {Romani}, {Rossi},
                        {Sch{\"o}nert}, {Semenov}, {Settanta}, {Skorokhvatov}, {Singhal}, {Smirnov},
                        {Sotnikov}, {Suvorov}, {Tartaglia}, {Testera}, {Thurn}, {Unzhakov},
                        {Villante}, {Vishneva}, {Vogelaar}, {von Feilitzsch}, {Wojcik}, {Wurm},
                        {Zavatarelli}, {Zuber}, \& {Zuzel}}]{Borexino-Collaboration20}
      {Borexino Collaboration}, Agostini, M., {Altenm{\"u}ller}, K., {Appel}, S.,
      {et~al.} 2020, \nat, 587, 577

      \bibitem[{{Buldgen} {et~al.}(2019){Buldgen}, {Salmon}, \&
                        {Noels}}]{Buldgen+19b}
      {Buldgen}, G., {Salmon}, S., \& {Noels}, A. 2019, Frontiers in Astronomy and
      Space Sciences, 6, 42

      \bibitem[{{Castro} {et~al.}(2007){Castro}, {Vauclair}, \&
                        {Richard}}]{Castro+07}
      {Castro}, M., {Vauclair}, S., \& {Richard}, O. 2007, \aap, 463, 755

      \bibitem[{{Christensen-Dalsgaard}(2021)}]{Christensen-Dalsgaard21}
      {Christensen-Dalsgaard}, J. 2021, Living Reviews in Solar Physics, 18, 2

      \bibitem[{{Christensen-Dalsgaard} {et~al.}(1996){Christensen-Dalsgaard},
                        {Dappen}, {Ajukov}, {Anderson}, {Antia}, {Basu}, {Baturin}, {Berthomieu},
                        {Chaboyer}, {Chitre}, {Cox}, {Demarque}, {Donatowicz}, {Dziembowski},
                        {Gabriel}, {Gough}, {Guenther}, {Guzik}, {Harvey}, {Hill}, {Houdek},
                        {Iglesias}, {Kosovichev}, {Leibacher}, {Morel}, {Proffitt}, {Provost},
                        {Reiter}, {Rhodes}, {Rogers}, {Roxburgh}, {Thompson}, \&
                        {Ulrich}}]{Christensen-Dalsgaard+96}
      {Christensen-Dalsgaard}, J., {Dappen}, W., {Ajukov}, S.~V., {et~al.} 1996,
      Science, 272, 1286

      \bibitem[{{Chugunov} {et~al.}(2007){Chugunov}, {Dewitt}, \&
                        {Yakovlev}}]{Chugunov+07}
      {Chugunov}, A.~I., {Dewitt}, H.~E., \& {Yakovlev}, D.~G. 2007, \prd, 76, 025028

      \bibitem[{Cox \& Giuli(1968)}]{Cox+Giuli68}
      Cox, J. \& Giuli, R. 1968, Gordon and Breach, New York, 401

      \bibitem[{{Eggenberger} {et~al.}(2022){Eggenberger}, {Buldgen}, {Salmon},
                        {Noels}, {Grevesse}, \& {Asplund}}]{Eggenberger+22}
      {Eggenberger}, P., {Buldgen}, G., {Salmon}, S.~J.~A.~J., {et~al.} 2022, Nature
      Astronomy, 6, 788

      \bibitem[{{Elbakyan} {et~al.}(2020){Elbakyan}, {Johansen}, {Lambrechts},
                        {Akimkin}, \& {Vorobyov}}]{Elbakyan+20}
      {Elbakyan}, V.~G., {Johansen}, A., {Lambrechts}, M., {Akimkin}, V., \&
      {Vorobyov}, E.~I. 2020, \aap, 637, A5

      \bibitem[{Farag {et~al.}(2020)Farag, Timmes, Taylor, Patton, \&
                        Farmer}]{Farag+20}
      Farag, E., Timmes, F.~X., Taylor, M., Patton, K.~M., \& Farmer, R. 2020, \apj,
      893, 133

      \bibitem[{{Gando} {et~al.}(2015){Gando}, {Gando}, {Hanakago}, {Ikeda}, {Inoue},
                        {Ishidoshiro}, {Ishikawa}, {Kishimoto}, {Koga}, {Matsuda}, {Matsuda},
                        {Mitsui}, {Motoki}, {Nakajima}, {Nakamura}, {Obata}, {Oki}, {Oki}, {Otani},
                        {Shimizu}, {Shirai}, {Suzuki}, {Tamae}, {Ueshima}, {Watanabe}, {Xu},
                        {Yamada}, {Yamauchi}, {Yoshida}, {Kozlov}, {Takemoto}, {Yoshida}, {Grant},
                        {Keefer}, {McKee}, {Piepke}, {Banks}, {Bloxham}, {Freedman}, {Fujikawa},
                        {Han}, {Hsu}, {Ichimura}, {Murayama}, {O'Donnell}, {Steiner}, {Winslow},
                        {Dwyer}, {Mauger}, {McKeown}, {Zhang}, {Berger}, {Lane}, {Maricic},
                        {Miletic}, {Learned}, {Sakai}, {Horton-Smith}, {Tang}, {Downum}, {Tolich},
                        {Efremenko}, {Kamyshkov}, {Perevozchikov}, {Karwowski}, {Markoff}, {Tornow},
                        {Detwiler}, {Enomoto}, {Heeger}, {Decowski}, \& {KamLAND
                                    Collaboration}}]{Gando+15}
      {Gando}, A., {Gando}, Y., {Hanakago}, H., {et~al.} 2015, \prc, 92, 055808

      \bibitem[{{Garaud}(2007)}]{Garaud07}
      {Garaud}, P. 2007, \apj, 671, 2091

      \bibitem[{{Gough}(2019)}]{Gough19}
      {Gough}, D.~O. 2019, \mnras, 485, L114

      \bibitem[{{Grevesse} \& {Sauval}(1998)}]{GS98}
      {Grevesse}, N. \& {Sauval}, A.~J. 1998, \ssr, 85, 161

      \bibitem[{Guillot \& Hueso(2006)}]{Guillot+Hueso06}
      Guillot, T. \& Hueso, R. 2006, \mnras, 367, L47

      \bibitem[{{Guillot} {et~al.}(2014){Guillot}, {Ida}, \& {Ormel}}]{Guillot+14}
      {Guillot}, T., {Ida}, S., \& {Ormel}, C.~W. 2014, \aap, 572, A72

      \bibitem[{{Guzik} {et~al.}(2005){Guzik}, {Watson}, \& {Cox}}]{Guzik+05}
      {Guzik}, J.~A., {Watson}, L.~S., \& {Cox}, A.~N. 2005, \apj, 627, 1049

      \bibitem[{{Guzik} {et~al.}(1987){Guzik}, {Willson}, \& {Brunish}}]{Guzik1987}
      {Guzik}, J.~A., {Willson}, L.~A., \& {Brunish}, W.~M. 1987, \apj, 319, 957

      \bibitem[{{Hartmann} {et~al.}(1998){Hartmann}, {Calvet}, {Gullbring}, \&
                        {D'Alessio}}]{Hartmann+98}
      {Hartmann}, L., {Calvet}, N., {Gullbring}, E., \& {D'Alessio}, P. 1998, \apj,
      495, 385

      \bibitem[{{Haxton} \& {Serenelli}(2008)}]{Haxton+Serenelli+08}
      {Haxton}, W.~C. \& {Serenelli}, A.~M. 2008, \apj, 687, 678

      \bibitem[{{Herwig}(2000)}]{Herwig00}
      {Herwig}, F. 2000, \aap, 360, 952

      \bibitem[{{Iben}(1965)}]{Iben65}
      {Iben}, Jr., I. 1965, \apj, 141, 993

      \bibitem[{{Iglesias} \& {Hansen}(2017)}]{Iglesias2017}
      {Iglesias}, C.~A. \& {Hansen}, S.~B. 2017, \apj, 835, 284

      \bibitem[{{Imbriani} {et~al.}(2005){Imbriani}, {Costantini}, {Formicola},
      {Vomiero}, {Angulo}, {Bemmerer}, {Bonetti}, {Broggini}, {Confortola},
      {Corvisiero}, {Cruz}, {Descouvemont}, {F{\"u}l{\"o}p}, {Gervino},
      {Guglielmetti}, {Gustavino}, {Gy{\"u}rky}, {Jesus}, {Junker}, {Klug},
      {Lemut}, {Menegazzo}, {Prati}, {Roca}, {Rolfs}, {Romano}, {Rossi-Alvarez},
      {Sch{\"u}mann}, {Sch{\"u}rmann}, {Somorjai}, {Straniero}, {Strieder},
      {Terrasi}, \& {Trautvetter}}]{Imbriani+05}
      {Imbriani}, G., {Costantini}, H., {Formicola}, A., {et~al.} 2005, European
      Physical Journal A, 25, 455

      \bibitem[{{Kippenhahn} \& {Weigert}(1990)}]{Kippenhahn+Weigert90}
      {Kippenhahn}, R. \& {Weigert}, A. 1990, {Stellar Structure and Evolution}
      (Springer-Verlag)

      \bibitem[{{Kruijer} {et~al.}(2020){Kruijer}, {Kleine}, \&
                        {Borg}}]{Kruijer+2020}
      {Kruijer}, T.~S., {Kleine}, T., \& {Borg}, L.~E. 2020, Nature Astronomy, 4, 32

      \bibitem[{{Kunitomo} \& {Guillot}(2021)}]{Kunitomo+Guillot21}
      {Kunitomo}, M. \& {Guillot}, T. 2021, \aap, 655, A51

      \bibitem[{{Kunitomo} {et~al.}(2018){Kunitomo}, {Guillot}, {Ida}, \&
                        {Takeuchi}}]{Kunitomo+18}
      {Kunitomo}, M., {Guillot}, T., {Ida}, S., \& {Takeuchi}, T. 2018, \aap, 618,
      A132

      \bibitem[{{Kunitomo} {et~al.}(2017){Kunitomo}, {Guillot}, {Takeuchi}, \&
                        {Ida}}]{Kunitomo+17}
      {Kunitomo}, M., {Guillot}, T., {Takeuchi}, T., \& {Ida}, S. 2017, \aap, 599,
      A49

      \bibitem[{{Kunitomo} {et~al.}(2020){Kunitomo}, {Suzuki}, \&
                        {Inutsuka}}]{Kunitomo+20}
      {Kunitomo}, M., {Suzuki}, T.~K., \& {Inutsuka}, S.-i. 2020, \mnras, 492, 3849

      \bibitem[{{Le Pennec} {et~al.}(2015){Le Pennec}, {Turck-Chi{\`e}ze}, {Salmon},
      {Blancard}, {Coss{\'e}}, {Faussurier}, \& {Mondet}}]{LePennec+15}
      {Le Pennec}, M., {Turck-Chi{\`e}ze}, S., {Salmon}, S., {et~al.} 2015, \apjl,
      813, L42

      \bibitem[{{Magg} {et~al.}(2022){Magg}, {Bergemann}, {Serenelli}, {Bautista},
                        {Plez}, {Heiter}, {Gerber}, {Ludwig}, {Basu}, {Ferguson}, {Gallego},
                        {Gamrath}, {Palmeri}, \& {Quinet}}]{Magg+22}
      {Magg}, E., {Bergemann}, M., {Serenelli}, A., {et~al.} 2022, \aap, 661, A140

      \bibitem[{{Manara} {et~al.}(2018){Manara}, {Morbidelli}, \&
                        {Guillot}}]{Manara+18}
      {Manara}, C.~F., {Morbidelli}, A., \& {Guillot}, T. 2018, \aap, 618, L3

      \bibitem[{{Montalban} {et~al.}(2006){Montalban}, {Miglio}, {Theado}, {Noels},
                        \& {Grevesse}}]{Montalban2006}
      {Montalban}, J., {Miglio}, A., {Theado}, S., {Noels}, A., \& {Grevesse}, N.
      2006, Communications in Asteroseismology, 147, 80

      \bibitem[{{Nahar} \& {Pradhan}(2016)}]{Nahar2016}
      {Nahar}, S.~N. \& {Pradhan}, A.~K. 2016, \prl, 116, 235003

      \bibitem[{Nelder \& Mead(1965)}]{Nelder+Mead65}
      Nelder, J.~A. \& Mead, R. 1965, The computer journal, 7, 308

      \bibitem[{Orebi~Gann {et~al.}(2021)Orebi~Gann, Zuber, Bemmerer, \&
      Serenelli}]{Orebi-Gann+21}
      Orebi~Gann, G.~D., Zuber, K., Bemmerer, D., \& Serenelli, A. 2021, Annual
      Review of Nuclear and Particle Science, 71, 491

      \bibitem[{{Particle Data Group} {et~al.}(2020){Particle Data Group}, {Zyla},
      {Barnett}, {Beringer}, {Dahl}, {Dwyer}, {Groom}, {Lin}, {Lugovsky},
      {Pianori}, {Robinson}, {Wohl}, {Yao}, {Agashe}, {Aielli}, {Allanach},
      {Amsler}, {Antonelli}, {Aschenauer}, {Asner}, {Baer}, {Banerjee}, {Baudis},
      {Bauer}, {Beatty}, {Belousov}, {Bethke}, {Bettini}, {Biebel}, {Black},
      {Blucher}, {Buchmuller}, {Burkert}, {Bychkov}, {Cahn}, {Carena}, {Ceccucci},
      {Cerri}, {Chakraborty}, {Chivukula}, {Cowan}, {D'Ambrosio}, {Damour}, {de
                  Florian}, {de Gouv{\^e}a}, {DeGrand}, {de Jong}, {Dissertori}, {Dobrescu},
      {D'Onofrio}, {Doser}, {Drees}, {Dreiner}, {Eerola}, {Egede}, {Eidelman},
      {Ellis}, {Erler}, {Ezhela}, {Fetscher}, {Fields}, {Foster}, {Freitas},
      {Gallagher}, {Garren}, {Gerber}, {Gerbier}, {Gershon}, {Gershtein},
      {Gherghetta}, {Godizov}, {Gonzalez-Garcia}, {Goodman}, {Grab}, {Gritsan},
      {Grojean}, {Gr{\"u}newald}, {Gurtu}, {Gutsche}, {Haber}, {Hanhart},
      {Hashimoto}, {Hayato}, {Hebecker}, {Heinemeyer}, {Heltsley},
      {Hern{\'a}ndez-Rey}, {Hikasa}, {Hisano}, {H{\"o}cker}, {Holder}, {Holtkamp},
      {Huston}, {Hyodo}, {Johnson}, {Kado}, {Karliner}, {Katz}, {Kenzie}, {Khoze},
      {Klein}, {Klempt}, {Kowalewski}, {Krauss}, {Kreps}, {Krusche}, {Kwon},
      {Lahav}, {Laiho}, {Lellouch}, {Lesgourgues}, {Liddle}, {Ligeti}, {Lippmann},
      {Liss}, {Littenberg}, {Lourengo}, {Lugovsky}, {Lusiani}, {Makida}, {Maltoni},
      {Mannel}, {Manohar}, {Marciano}, {Masoni}, {Matthews}, {Mei{\ss}ner},
      {Mikhasenko}, {Miller}, {Milstead}, {Mitchell}, {M{\"o}nig}, {Molaro},
      {Moortgat}, {Moskovic}, {Nakamura}, {Narain}, {Nason}, {Navas}, {Neubert},
      {Nevski}, {Nir}, {Olive}, {Patrignani}, {Peacock}, {Petcov}, {Petrov},
      {Pich}, {Piepke}, {Pomarol}, {Profumo}, {Quadt}, {Rabbertz}, {Rademacker},
      {Raffelt}, {Ramani}, {Ramsey-Musolf}, {Ratcliff}, {Richardson}, {Ringwald},
      {Roesler}, {Rolli}, {Romaniouk}, {Rosenberg}, {Rosner}, {Rybka}, {Ryskin},
      {Ryutin}, {Sakai}, {Salam}, {Sarkar}, {Sauli}, {Schneider}, {Scholberg},
      {Schwartz}, {Schwiening}, {Scott}, {Sharma}, {Sharpe}, {Shutt}, {Silari},
      {Sj{\"o}strand}, {Skands}, {Skwarnicki}, {Smoot}, {Soffer}, {Sozzi},
      {Spanier}, {Spiering}, {Stahl}, {Stone}, {Sumino}, {Sumiyoshi}, {Syphers},
      {Takahashi}, {Tanabashi}, {Tanaka}, {Ta{\v{s}}evsk{\'y}}, {Terashi},
      {Terning}, {Thoma}, {Thorne}, {Tiator}, {Titov}, {Tkachenko}, {Tovey},
      {Trabelsi}, {Urquijo}, {Valencia}, {Van de Water}, {Varelas}, {Venanzoni},
      {Verde}, {Vincter}, {Vogel}, {Vogelsang}, {Vogt}, {Vorobyev}, {Wakely},
      {Walkowiak}, {Walter}, {Wands}, {Wascko}, {Weinberg}, {Weinberg}, {White},
      {Wiencke}, {Willocq}, {Woody}, {Workman}, {Yokoyama}, {Yoshida},
      {Zanderighi}, {Zeller}, {Zenin}, {Zhu}, {Zhu}, {Zimmermann}, {Anderson},
      {Basaglia}, {Lugovsky}, {Schaffner}, \& {Zheng}}]{Zyla+20}
      {Particle Data Group}, {Zyla}, P.~A., {Barnett}, R.~M., {et~al.} 2020, Progress
      of Theoretical and Experimental Physics, 2020, 083C01

      \bibitem[{{Paxton} {et~al.}(2011){Paxton}, {Bildsten}, {Dotter}, {Herwig},
                        {Lesaffre}, \& {Timmes}}]{Paxton+11}
      {Paxton}, B., {Bildsten}, L., {Dotter}, A., {et~al.} 2011, \apjs, 192, 3

      \bibitem[{{Paxton} {et~al.}(2013){Paxton}, {Cantiello}, {Arras}, {Bildsten},
                        {Brown}, {Dotter}, {Mankovich}, {Montgomery}, {Stello}, {Timmes}, \&
                        {Townsend}}]{Paxton+13}
      {Paxton}, B., {Cantiello}, M., {Arras}, P., {et~al.} 2013, \apjs, 208, 4

      \bibitem[{{Paxton} {et~al.}(2015){Paxton}, {Marchant}, {Schwab}, {Bauer},
                        {Bildsten}, {Cantiello}, {Dessart}, {Farmer}, {Hu}, {Langer}, {Townsend},
                        {Townsley}, \& {Timmes}}]{Paxton+15}
      {Paxton}, B., {Marchant}, P., {Schwab}, J., {et~al.} 2015, \apjs, 220, 15

      \bibitem[{{Paxton} {et~al.}(2018){Paxton}, {Schwab}, {Bauer}, {Bildsten},
                        {Blinnikov}, {Duffell}, {Farmer}, {Goldberg}, {Marchant}, {Sorokina},
                        {Thoul}, {Townsend}, \& {Timmes}}]{Paxton+18}
      {Paxton}, B., {Schwab}, J., {Bauer}, E.~B., {et~al.} 2018, \apjs, 234, 34

      \bibitem[{{Paxton} {et~al.}(2019){Paxton}, {Smolec}, {Schwab}, {Gautschy},
                        {Bildsten}, {Cantiello}, {Dotter}, {Farmer}, {Goldberg}, {Jermyn}, {Kanbur},
                        {Marchant}, {Thoul}, {Townsend}, {Wolf}, {Zhang}, \& {Timmes}}]{Paxton+19}
      {Paxton}, B., {Smolec}, R., {Schwab}, J., {et~al.} 2019, \apjs, 243, 10

      \bibitem[{{Sackmann} \& {Boothroyd}(2003)}]{Sackmann+Boothroyd03}
      {Sackmann}, I.~J. \& {Boothroyd}, A.~I. 2003, \apj, 583, 1024

      \bibitem[{{Salmon} {et~al.}(2021){Salmon}, {Buldgen}, {Noels}, {Eggenberger},
                        {Scuflaire}, \& {Meynet}}]{Salmon+21}
      {Salmon}, S.~J.~A.~J., {Buldgen}, G., {Noels}, A., {et~al.} 2021, \aap, 651,
      A106

      \bibitem[{{Sato} {et~al.}(2016){Sato}, {Okuzumi}, \& {Ida}}]{Sato+16}
      {Sato}, T., {Okuzumi}, S., \& {Ida}, S. 2016, \aap, 589, A15

      \bibitem[{{Serenelli} {et~al.}(2011){Serenelli}, {Haxton}, \&
                        {Pe{\~n}a-Garay}}]{Serenelli+11}
      {Serenelli}, A.~M., {Haxton}, W.~C., \& {Pe{\~n}a-Garay}, C. 2011, \apj, 743,
      24

      \bibitem[{{Thoul} {et~al.}(1994){Thoul}, {Bahcall}, \& {Loeb}}]{Thoul+94}
      {Thoul}, A.~A., {Bahcall}, J.~N., \& {Loeb}, A. 1994, \apj, 421, 828

      \bibitem[{{Tsukamoto} {et~al.}(2017){Tsukamoto}, {Okuzumi}, \&
                        {Kataoka}}]{Tsukamoto+17}
      {Tsukamoto}, Y., {Okuzumi}, S., \& {Kataoka}, A. 2017, \apj, 838, 151

      \bibitem[{{Vescovi} {et~al.}(2019){Vescovi}, {Piersanti}, {Cristallo}, {Busso},
                        {Vissani}, {Palmerini}, {Simonucci}, \& {Taioli}}]{Vescovi+19}
      {Vescovi}, D., {Piersanti}, L., {Cristallo}, S., {et~al.} 2019, \aap, 623, A126

      \bibitem[{{Villante} \& {Serenelli}(2021)}]{Villante+Serenelli21}
      {Villante}, F.~L. \& {Serenelli}, A. 2021, Frontiers in Astronomy and Space
      Sciences, 7, 112

      \bibitem[{{Vinyoles} {et~al.}(2017){Vinyoles}, {Serenelli}, {Villante}, {Basu},
                        {Bergstr{\"o}m}, {Gonzalez-Garcia}, {Maltoni}, {Pe{\~n}a-Garay}, \&
                        {Song}}]{Vinyoles+17}
      {Vinyoles}, N., {Serenelli}, A.~M., {Villante}, F.~L., {et~al.} 2017, \apj,
      835, 202

      \bibitem[{{Wood} {et~al.}(2005){Wood}, {M{\"u}ller}, {Zank}, {Linsky}, \&
                        {Redfield}}]{Wood+05}
      {Wood}, B.~E., {M{\"u}ller}, H.-R., {Zank}, G.~P., {Linsky}, J.~L., \&
      {Redfield}, S. 2005, \apjl, 628, L143

      \bibitem[{{Zhang} {et~al.}(2019){Zhang}, {Li}, \&
                        {Christensen-Dalsgaard}}]{Zhang+19}
      {Zhang}, Q.-S., {Li}, Y., \& {Christensen-Dalsgaard}, J. 2019, \apj, 881, 103

\end{thebibliography}
\end{document}